\documentclass[11pt]{article}
\pdfoutput=1

\usepackage{comment}
\usepackage{framed}
\usepackage{psfrag}
\usepackage{array}
\usepackage{bm, bbm, bbold}
\usepackage{amsmath,amsfonts,amssymb, mathrsfs, amsthm}
\usepackage{braket}
\usepackage{graphicx}
\usepackage[labelsep=quad]{subcaption}
\usepackage{color,soul}
\usepackage[punctsep]{collref}
\collectsep[]{;}
\usepackage{epsfig}
\usepackage{wrapfig}
\usepackage{tikz}
\usepackage{gensymb}
\usetikzlibrary{arrows,plotmarks,positioning,calc}
\usepackage{environ}
\NewEnviron{eqn}{
\begin{align}
\begin{split}
  \BODY
\end{split}
\end{align}
}
\NewEnviron{eqn*}{
\begin{align*}
\begin{split}
  \BODY
\end{split}
\end{align*}
}
\newcommand{\be}{\begin{equation}}
\newcommand{\ee}{\end{equation}}
\newcommand{\bes}{\begin{equation*}}
\newcommand{\ees}{\end{equation*}}
\newcommand{\nno}{\nonumber}

\usepackage{float}

\newcommand{\largeeq}[1]{\text{\large$#1$}}

\usepackage{tikz} 
\usetikzlibrary{decorations.pathreplacing}

\usepackage{jheppub}
\usepackage{hyperref}

\usetikzlibrary{decorations.pathmorphing}	
\usetikzlibrary{decorations.markings}
\tikzset{
    sugra/.style={decorate, decoration={snake}, draw=black},
    scalarphi/.style={dashed,draw=black, postaction={decorate},
        },
    hwbou/.style={draw=blue, postaction={decorate}, ultra thick
        },
    vector/.style={draw=blue,decorate, decoration={snake}, draw},
	provector/.style={decorate, decoration={snake,amplitude=2.5pt}, draw},
	antivector/.style={decorate, decoration={snake,amplitude=-2.5pt}, draw},
   	 fermion/.style={draw=cyan, postaction={decorate},
        decoration={markings,mark=at position .55 with {\arrow[draw=black]{>}}}},
    fermionbar/.style={draw=cyan, postaction={decorate},
        decoration={markings,mark=at position .55 with {\arrow[draw=black]{<}}}},
    fermionnoarrow/.style={draw=black},
    gluon/.style={decorate, draw=red,
        decoration={coil,amplitude=4pt, segment length=5pt}},
    scalar/.style={dashed,draw=black, postaction={decorate},
        decoration={markings,mark=at position .55 with {\arrow[draw=black]{>}}}},
    scalarbar/.style={dashed,draw=black, postaction={decorate},
        decoration={markings,mark=at position .55 with {\arrow[draw=black]{<}}}},
    electron/.style={draw=black, postaction={decorate},
        decoration={markings,mark=at position .55 with {\arrow[draw=black]{>}}}},
    scalarnoarrow/.style={dashed, draw=black},
    electron/.style={draw=black, postaction={decorate},
        decoration={markings, mark=at position .55 with {\arrow[draw=black]{>}}}},
	bigvector/.style={decorate, decoration={snake, amplitude=4pt}, draw},
    photon/.style={draw=red, decorate, decoration={zigzag}, draw},
    higgs/.style={dashed, draw=black, postaction={decorate},
        },	
        goldstone/.style={draw=brown, postaction={decorate},
        },    
          ghost/.style={dashed, draw=magenta, postaction={decorate},
        decoration={markings, mark=at position .55 with {\arrow[draw=black]{>}}}
        },  
          antighost/.style={dashed, draw=magenta, postaction={decorate},
        decoration={markings, mark=at position .55 with {\arrow[draw=black]{<}}}
        },  
          mphoton/.style={decorate, decoration={snake}, draw=violet},
            realscalar/.style={draw=black}, 
           mgluon/.style={decorate, draw=blue,
        decoration={coil,amplitude=4pt, segment length=5pt}},
         weylfermion/.style={draw=orange, postaction={decorate},
        decoration={markings,mark=at position .55 with {\arrow[draw=black]{>}}}},
         weylfermionbar/.style={draw=orange, postaction={decorate},
        decoration={markings,mark=at position .55 with {\arrow[draw=black]{<}}}}, 
   	wboson/.style={draw=blue,decorate, decoration={snake,amplitude=4pt}, draw},  
    zboson/.style={draw=violet, decorate, decoration={snake}, draw},   
    lepton/.style={draw=black, postaction={decorate},
        decoration={markings,mark=at position .55 with {\arrow[draw=black]{>}}}},
    leptonbar/.style={draw=black, postaction={decorate},
        decoration={markings,mark=at position .55 with {\arrow[draw=black]{<}}}}, 
        graviton/.style={draw=blue,decorate, decoration={snake,amplitude=4pt}, draw},  
        gravitino/.style={draw=red, postaction={decorate},
        decoration={snake, markings, mark=at position .55 with {\arrow[draw=black]{>}}}},
    gravitinobar/.style={draw=red, postaction={decorate},
        decoration={snake, markings,mark=at position .55 with {\arrow[draw=black]{<}}} },    
}

\def\e{\epsilon}
\def\q{\theta}

\def\p{\partial}

\newcommand{\intsinf}{\int_{0}^{\infty}} 
\newcommand{\intnsinf}{\int_{-\infty}^{0}}

\begin{document}    
\title{Analytic Results for Loop-Level Momentum Space Witten Diagrams}
\author{Chandramouli Chowdhury$^{1 }$ and Kajal Singh$^{2, 3}$ }
\emailAdd{chandramouli.chowdhury@icts.res.in, kajalsingh@hri.res.in}

\affiliation{$^1$International Centre for Theoretical Sciences, Tata Institute of Fundamental Research, Shivakote,
Bengaluru 560089, India.\\
$^{2}$Harish-Chandra Research Institute,  Chhatnag Road, Jhunsi, Allahabad 211019,
India.\\
$^{3}$Homi Bhabha National Institute, Training School Complex, Anushaktinagar,
Mumbai 400094, India}

\abstract{This paper presents an evaluation of the wave function coefficients for conformally coupled scalars at both one and two-loop levels at leading order in the coupling constant, in momentum space. We take cues from time-dependent interactions in flat spacetime and under suitable approximations, these can also be used to study the wave function coefficients for general cosmologies. We make use of recursion relations developed in arxiv:\{1709.02813\} to regularize certain bulk-point integrals and express the wave function coefficients in a form that simplifies the loop integrals. We utilize hard-cutoff regularization to regularize the loop integrals and further provide a discussion on their renormalization. Our results can also be analytically continued to obtain answers for transition amplitudes in AdS. 
}

\maketitle
\noindent
\flushbottom
\section{Introduction}
 Scattering amplitudes are one of the most basic objects in the subject of quantum field theory (QFT) and quantum gravity (QG). They have received significant attention in the last couple of decades, and this attention has led to the discovery of several hidden structures (for example, the discovery of a geometrical object known as the Amplituhedron \cite{Arkani-Hamed:2013jha}). One of the most widely used methods for studying scattering amplitudes is perturbation theory. The conventional approach to calculating these amplitudes is through the use of Feynman diagrams, which provide a diagrammatic representation of the terms in perturbation theory. Although these diagrams make locality and Lorentz invariance of a theory manifest, they are not the best method to compute scattering amplitudes involving large numbers of particles.
This is because the computation of such amplitudes involves summing over a multitude of Feynman diagrams with various tensor structures and it is not feasible to extract any physical insight unless the answers can be expressed in a concise form.
 A notable example of this is in the amplitude which showed up in the study of scattering of gluons. It was found by Parke and Taylor \cite{Parke:1986gb} that the six-point amplitude for the scattering of gluons had a very compact expression despite having many Feynman diagrams showing up in the intermediate stages of the computation. This is now known as the Parke-Taylor formula.  Due to the simplicity of the final answer, it was anticipated that there existed an easier and alternative approach for computing it, which did not require the use of Feynman diagrams.  This easier way was later discovered in papers by Britto, Cachazo, Feng, and Witten \cite{Britto:2004ap, Cachazo:2004kj, Britto:2005fq} and are known as the BCFW recursion relations. Significant advancements have been made in this subject following the seminal works of BCFW. 

 There has been a renewed interest in the study of cosmological correlation functions since the work of Maldacena \cite{Maldacena:2002vr}.  Studying these correlation functions enables cosmologists to gain knowledge about various phenomena like galaxy formation, dark energy, and the early history of the universe. They are also related to quantities that can be measured in experiments by studying fluctuations in the cosmic microwave background (CMB) radiation. One of the active areas of interest has been to study the non-Gaussianities in the CMB. These refer to fluctuations that deviate from the Gaussian normal distribution of the density fluctuations in the CMB. They are predicted by simple models of inflation, where they are generated by quantum fluctuations of the inflaton field during the inflationary epoch. To theoretically compute cosmological correlation functions, a more comprehensive understanding of the state of our universe is required. This state, typically defined by a path integral, is commonly referred to as the \textit{wave function of the universe} \cite{Hartle:1983ai} and can be computed in perturbation theory. The AdS/CFT correspondence \cite{Maldacena:1997re, Witten:1998qj} establishes a connection between the wave function of the universe in dS space and transition amplitudes in AdS, which we shall comment on towards the end.

 The wave function of the universe can be computed in various ways in perturbation theory. There have been several approaches to studying this, for example, in position space \cite{Bertan:2018khc,  Heckelbacher:2022fbx, Carmi:2019ocp, Carmi:2021dsn}, Mellin space \cite{Fitzpatrick:2011ia, Sleight:2019mgd}, etc. There has also been significant progress in studying these in momentum space \cite{Raju:2010by, Raju:2011mp, Raju:2012zs, Arkani-Hamed:2015bza, Arkani-Hamed:2017fdk, Benincasa:2022gtd, Konstantinidis:2016nio, Costantino:2020vdu} including several recursion relations that have simplified the computations at tree level, e.g., BCFW recursion relations \cite{Raju:2010by}, ``clipping rules'' for scalars \cite{Arkani-Hamed:2017fdk} and gluons \cite{Albayrak:2019asr}, differential equations for correlation functions \cite{Arkani-Hamed:2018kmz}.  In particular, the paper \cite{Arkani-Hamed:2017fdk} also introduced new geometrical objects known as the \textit{Cosmological Polytopes} whose canonical form was related to the wave function of the universe. The discovery of this mathematical object aligns with the general theme of relating geometry with scattering amplitudes \cite{Arkani-Hamed:2013jha, Arkani-Hamed:2017tmz, Arkani-Hamed:2017mur, Arkani-Hamed:2020blm}. This is not surprising as the wave function of the universe secretly contains the flat space scattering amplitude, in what is known as the flat space limit \cite{Raju:2012zr}. One can gain information about Lorentz invariance, unitarity, and other aspects of the theory through different triangulations and facets of the cosmological polytope \cite{Arkani-Hamed:2017fdk, Arkani-Hamed:2018bjr}. While there has been considerable progress in calculating the wave function at the tree level, advancements in computing it at the loop level have been limited \cite{Carmi:2019ocp, Gorbenko:2019rza, Albayrak:2020isk, Carmi:2021dsn, Banados:2022nhj, Fichet:2021xfn}.  In the paper \cite{Albayrak:2020isk}, the authors used the recursion relations developed in \cite{Arkani-Hamed:2017fdk} to evaluate the value of transition amplitudes for conformally coupled scalars in AdS in momentum space. These were evaluated for some special cases of bubble diagrams at one-loop for conformally coupled interactions in four-dimensions. 

In this paper we evaluate one and two-loop diagrams for conformally coupled scalar in four and six dimensions. In particular, we compute all possible diagrams at two-loop for the 2-point function (which, as explained later, is equivalent to the wave function coefficient at the second order) in the $\phi^4$ theory in dS$_4$.  We also find a way to regularize certain classes of Witten diagrams, like the cactus diagram (see figure \ref{fig:2cactus}) which are naively divergent even at the level of the integrand. We also compute the one loop triangle diagram for $\phi^3$ theory in dS$_6$ in the squeezed limit and then discuss how one can extract the general answer. We also find that the finite part of the diagrams have the same behavior as that of the 2-point and 3-point functions in a CFT living in one lower dimension.  We demonstrate how these methods can be used to evaluate higher point diagrams for more general theories, such as the 4-point bubble diagram in dS$_4$ for $g \phi^3 + \lambda \phi^4 $ theory. These shed light on the analytic structure of loop-level Witten diagrams for conformally coupled scalars. Although our central results are primarily derived for interactions that are conformally invariant, toward the end, we also comment on the structure of the wave function for other kinds of couplings.

\section{Review of Perturbation Theory and Recursion Relations}\label{sec:review}
In this section, we review the construction of \cite{Arkani-Hamed:2017fdk} and introduce some of the building blocks for computing Witten diagrams. We will keep our discussion self-contained and encourage readers who seek additional details to refer to \cite{Arkani-Hamed:2017fdk} and the review \cite{Benincasa:2022gtd}.

We will start with a self-interacting massless scalar field in FLRW$_{d+1}$ with the action given as 
\be\label{action1}
 S = -  \int d^{d}x d\eta\, \sqrt{-g} \left[  \frac{1}{2} g^{\mu\nu} \p_\mu \phi \p_\nu \phi +\xi R \phi^2 + \sum_{n \geq 3} \frac{\lambda_n}{n!} \phi^n \right]
\ee
where $g_{\mu\nu}$ is the metric of FLRW$_{d+1}$ in the Poincare patch and is given as 
\be
ds^2 = a^2(\eta)\big[- d\eta^2 + \delta_{ij} dx^i dx^j\Big]
\ee
where the index $i, j$ runs over the $d$-spatial directions and $\eta$ is the conformal time. Note that we have set the Hubble parameter $H = 1$. The time-like boundary of spacetime is located at $\eta\to 0$. The scalar fields have a scaling dimension of $\Delta = \frac{d-1}{2}$.

In this paper,  we draw insights by working with conformally coupled scalars where the interaction terms are also conformally invariant. We later generalize some of our  results to other interactions. 
The advantage of working with conformally coupled scalars is that we can use a Weyl transformation $\phi \to a^{- \Delta}(\eta) \phi$ and $g_{\mu\nu} \to a^2(\eta) g_{\mu\nu}$ to map the problem in FLRW to that of flat spacetime with massless scalars. This requirement sets $\xi = \frac{d-1}{4d}$ and we end up with the following action 
\bes
S =  \int d^dxd\eta \, \Big(\frac{1}{2} (\p_\eta \phi)^2 - \frac{1}{2}  (\p_i \phi)^2 - \sum_{n \geq 3} [a(\eta)]^{(2 - n)\Delta +2}\frac{\lambda_n \phi^n }{n!}  \Big)~.
\ees
By absorbing $[a(\eta)]^{(2 - k)\Delta +2}$ in the definition of $\lambda$ we effectively have the Lagrangian for a massless scalar field in flat space with time-dependent interactions \footnote{By taking loop corrections into account, it may not be possible to invert this conformal transformation and obtain results for dS space. It would be interesting to determine where the modification needs to be made in order to evaluate the result in dS space. However, for time-dependent interactions in flat space, this setup remains perfectly valid.},
\be\label{action2}
S =  \int d^dxd\eta  \,\Big(\frac{1}{2} (\p_\eta \phi)^2 - \frac{1}{2}  (\p_i \phi)^2 -\sum_{n \geq 3} \frac{\lambda_n(\eta) \phi^n }{n!}  \Big)
\ee
where
\be\label{coupling}
\lambda_n(\eta) = [a(\eta)]^{(2 - n)\Delta +2} \lambda_n~.
\ee

For the interactions to be time-independent (which is equivalent to them being conformally invariant), the parameter $n$ takes a specific value depending on the dimension of spacetime \cite{Albayrak:2020isk},
\be\label{conf-choice}
n =2 \frac{\Delta+1}{\Delta} = \frac{2(d+1)}{d-1}~.
\ee
This shows that there are three possible choices for $(n, D \equiv d+1)$ such that $n$ and $D$ are positive integers. These are $(6, 3), \ (4, 4), \ (3, 6)$. Since $D= 3$ is not a favorable spacetime to study the propagation of fields with general spin, we shall restrict to the case of $D=4$ and $D=6$. Thus, the only cases we are left with are
$\phi^4$ theory in $D=4$ and $\phi^3$ theory in $D=6$. We will later relax this restriction and consider more general cases.  

The wave function of the universe is defined by the following path integral 
\be\label{PI-defn}
\Psi_U[\phi(\bm x)] = \int_{\varphi(-\infty(1 - i \e), \bm x) = 0}^{\varphi(0, \bm x) = \phi(\bm x)} D\varphi \, e^{i S[\varphi]}~,
\ee
and it can be normalized using $\int D\phi \,|\Psi_U[\phi(x)]|^2 = 1$~. 
This path integral can be evaluated using Saddle point methods and it is convenient to go to momentum space and Fourier transform $\varphi(\eta, \bm x) = \int e^{i \bm k \cdot \bm x} \varphi(\eta, \bm k) d^d x$. For evaluating this integral using this method it is also convenient to split the field $\varphi(\eta, \bm k)$ into a ``classical'' and a ``quantum part'',  
\be\label{splitfields}
\varphi(\eta, \bm k) = \varphi_{cl}(\eta, \bm k) + \delta \varphi(\eta, \bm k)
\ee
where $\varphi_{cl}(\eta, \bm x)$ satisfies the  equation of motion for to the action given in \eqref{action2},
\be
\p_{\eta}^2\varphi_{cl}(\eta, \bm k) - k^2 \varphi_{cl}(\eta, \bm k) + \frac{\lambda_k(\eta)}{(k-1)!}  \varphi^{k-1}_{cl}(\eta, \bm k) = 0~.
\ee
and the boundary conditions at $\eta \to -\infty$ and $\eta \to 0$~. This equation is non-linear and can be solved using the Green's function approach. The general solution can be written down as 
\be\label{cl-soln}
\varphi_{cl}(\eta, \bm k)= \phi(x) e^{+ i \eta |\bm k|} - i \int d\eta'\, G(\eta, \eta', \bm k)  \frac{\lambda_k(\eta)}{(k-1)!}  \varphi^{k-1}_{cl}(\eta, \bm k)~,
\ee
The first term in this is the solution of the free equation of motion $(\lambda_k(\eta) = 0)$~, which are in general plane waves of the kind $e^{\pm i |\bm k|\eta}$. The boundary condition at $\eta \to -\infty$ only allows for one sign $ e^{+i|\bm k| \eta} $ and this choice is commonly referred to as the Bunch-Davies vacuum. This gives the bulk-boundary propagator\footnote{This often requires a normalization to account for the IR divergence as one goes towards $\eta\to0$. Therefore each of these bulk-boundary propagators comes with a factor of $\frac{1}{\eta_* |\bm k|}$ with $\eta_* \to 0$ defining the IR regulator. Since  these are overall factors, we shall suppress them in these computations.},
\be\label{bulk-bndy}
B(\eta, \bm k) = e^{i |\bm k|\eta}~.
\ee
The second term in \eqref{cl-soln} introduces the \textit{bulk-bulk propagator} and it is given as the following Green's function (where $k \equiv |\bm k|$),
\be\label{bulkbulkdS}
G(\eta, \eta', k) = \frac{1}{2k}\big[ e^{- 
i k (\eta - \eta')} \Theta(\eta - \eta') + e^{- 
i k (\eta' - \eta)} \Theta(\eta' - \eta) - e^{i k (\eta + \eta')} \big]~.
\ee
Using \eqref{splitfields} one can integrate out $\varphi_{cl}$, then performing the path integral over $\delta\varphi$ perturbatively gives us an expansion analogous to the usual Feynman diagrammatic expansion. The difference between this and the usual diagrammatic expansion in flat space is that we now have two kinds of propagators: the bulk-bulk propagator \eqref{bulkbulkdS} and bulk-boundary propagator \eqref{bulk-bndy}. The final wave function then takes the following form 
\be
\Psi_U = \exp \left\{ \sum_{n \geq 2} \frac{1}{n!} \int \prod_{v = 1}^n d^d x_v\, \phi(z_v) \delta\Bigl(\sum_n \bm k_n\Bigr)\tilde \psi_n(z) \right\}~,
\ee
where $\tilde\psi_n$ (which shall be referred to as the \textit{wave function coefficients}) are represented by integrals\footnote{The dependence of the lower limit on $i\e$ has been suppressed from this integral.}
\be\label{wavefunccoeff1}
\tilde \psi_n(E_v, E_e) = \int_{-\infty}^0 \prod_{v, e} d\eta_v\, e^{i \eta_v x_v} G(\eta_v, \eta_{v'}, y_e)~,
\ee
where $x_v$ denotes the sum of energies $\sum_a |\bm k_a| $ at a vertex and $y$ denotes the energy\footnote{The word ``energy'' is being used loosely here as there is no strict notion of on-shellness. Whenever we use the word energy, we always mean the modulus of momenta.} of an internal line, e.g., $|\bm k_1 + \bm k_2|$~. The integrals can be arranged in the form of diagrammatic representations, which are referred to as \textit{Witten diagrams}. These diagrams are analogous to the typical Feynman diagrams that are utilized in S-matrix calculations. In section \ref{sec:review}, we provide a detailed examination of the expansion of wave function coefficients in terms of Witten diagrams. 

For theories where $\lambda_k$ is time-dependent (i.e, the interactions are not conformally coupled) the wave function \eqref{wavefunccoeff1} takes the form \cite{Arkani-Hamed:2017fdk},
\be\label{wavefunccoeff2}
\tilde \psi_n(E_v, E_e) = \intnsinf \prod_v d\e_v \prod_{v \in \mathcal V} \tilde \lambda_{k_v}(\e) \psi(E_v + \e_v, E_e)~,
\ee
where $\lambda_k(\eta) = \intnsinf d\e\, e^{i \e \eta} \tilde \lambda_k(\e)$~. 
Thus, for computing the wave function coefficient for time-dependent $\lambda(\eta)$, we need to perform additional integrals on top of the wave function coefficient for the conformally coupled case. 
For the majority of this paper, we will analyze cases involving conformally coupled self-interactions, where $\lambda_k(\eta)$ is equivalent to a constant $\lambda_k$. In section \ref{sec:Concl}, we will demonstrate how these answers can be applied to derive results for more general cases.

We now summarize the diagrammatic representations for the wave function coefficients given in equation \eqref{wavefunccoeff1}. Consider the leading order terms in perturbation theory for three and four-point functions in $\phi^3$ theory. These correspond to the following Witten diagrams,
\begin{figure}[H]
\centering
\begin{tikzpicture}
\begin{scope}[shift={(-4, 0)}]
 \draw[thick, black] (-2, 0) -- (2, 0); 
 \draw[fermion, black] (-1.5, 0) -- (0, -2.5);
 \draw[fermion, black] (1.5, 0) -- (0, -2.5);
 \draw[fermion, black] (0, 0) -- (0, -2.5);
 \node at (-1.25, -1) {$\bm k_1$};
 \node at (-0.25, -1) {$\bm k_2$};
\node at (1.25, -1) {$\bm k_3$};
\end{scope}

\begin{scope}[shift={(4, 0)}]
 \draw[thick, black] (-3, 0) -- (3, 0); 
 \draw[fermion, black] (-2, 0) -- (-1.5, -2.5);
 \draw[fermion, black] (2, 0) -- (1.5, -2.5);
 \draw[fermion, black] (-1, 0) -- (-1.5, -2.5);
 \draw[fermion, black] (1, 0) -- (1.5, -2.5);
 \node at (-2.1, -1) {$\bm k_1$};
 \node at (-0.8, -1) {$\bm k_2$};
\node at (0.8, -1) {$\bm k_3$};
\node at (2.1, -1) {$\bm k_4$};
\draw[fermion, black] (-1.5, -2.5) -- (1.5, -2.5);
\node at (0, -2.75) {$\bm k_1 + \bm k_2$};
\end{scope}
\end{tikzpicture}
\caption{
 The boundary, $\eta \to 0$, of spacetime is represented by the  bold horizontal line and the edges starting from that line are the bulk-boundary propagators \eqref{bulk-bndy} and the edge in the bulk (of the diagram on the right) is the bulk-bulk propagator \eqref{bulkbulkdS}.}
 \label{WittenEX1}
\end{figure}
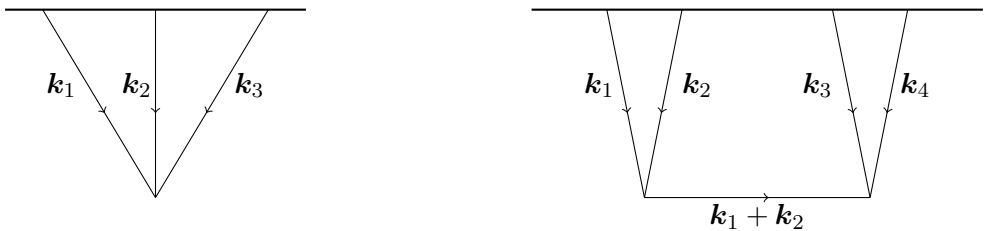
Consider the diagram on the left in figure \ref{WittenEX1}. This diagram is given by integrating over the bulk point, $\eta$, and attaching the three bulk-boundary propagators\footnote{We also use the notation $k_i \equiv |\bm k_i|$.} \footnote{Note that we are suppressing all the $i \e$ as our integrals are finite without using them. The $i \e$ prescription comes naturally while performing the $\eta$ integrals as the lower limit of the integral has $-\infty(1 - i \e)$. It is possible to restore these $i \e$  factors and perform the integrals. It would be interesting to see if different $i\e$ prescriptions allow us to obtain different correlation functions. }
\begin{eqn}\label{tree1}
\begin{tikzpicture}[baseline]
\draw[thick, black] (-2, 1.5) -- (2, 1.5); 
 \draw[fermion, black] (-1.5, 1.5) -- (0, -1.5);
 \draw[fermion, black] (1.5, 1.5) -- (0, -1.5);
 \draw[fermion, black] (0, 1.5) -- (0, -1.5);
 \node at (-1.25, 0.5) {$\bm k_1$};
 \node at (-0.25, 0.5) {$\bm k_2$};
\node at (1.25, 0.5) {$\bm k_3$};
\node at (0, -1.75) {$\eta$};
\end{tikzpicture}
= 
\intnsinf e^{i (k_1 + k_2 + k_3)\eta} d\eta = \frac{-i}{k_1 + k_2 + k_3} ~.
\end{eqn}
The diagram on the right in fig \ref{WittenEX1} can be expressed in a similar manner, 
\begin{eqn}\label{tree2}
    &\begin{tikzpicture}[baseline]
   \draw[thick, black] (-3, 1.5) -- (3, 1.5); 
 \draw[fermion, black] (-2, 1.5) -- (-1.5, -1);
 \draw[fermion, black] (2, 1.5) -- (1.5, -1);
 \draw[fermion, black] (-1, 1.5) -- (-1.5, -1);
 \draw[fermion, black] (1, 1.5) -- (1.5, -1);
 \node at (-2.1, 0.5) {$\bm k_1$};
 \node at (-0.8, 0.5) {$\bm k_2$};
\node at (0.8, 0.5) {$\bm k_3$};
\node at (2.1, 0.5) {$\bm k_4$};
\draw[fermion, black] (-1.5, -1) -- (1.5, -1);
\node at (0, -1.25) {$\bm k_1 + \bm k_2$};
\node at (-1.5, -1.25) {$\eta_1$};
\node at (1.5, -1.25) {$\eta_2$};
    \end{tikzpicture}
    = \intnsinf d\eta_1 d\eta_2 e^{i (k_1 + k_2) \eta_1} e^{i (k_3 + k_4)\eta_2} G(\eta_1, \eta_2, |\bm k_1 + \bm k_2|) \\
    &= \intnsinf d\eta_1 d\eta_2 e^{i (k_1 + k_2) \eta_1} e^{i (k_3 + k_4)\eta_2} \frac{1}{2k_I} \big[ \Theta(\eta_1 - \eta_2) e^{- i k_I (\eta_1 - \eta_2)} +  \Theta(\eta_2 - \eta_1) e^{- i k_I (\eta_2 - \eta_1)} - e^{i k_I (\eta_1 + \eta_2)} \big]~,
\end{eqn}
where we denote $k_I = |\bm k_1 + \bm k_2|$. 
We are not going to evaluate the integral explicitly   now but will give an iterative procedure to do so in the next section. 

We now introduce some more notations that we shall be using. The bold horizontal line depicting $\eta \to 0$ shall be removed altogether and bulk-boundary propagators shall be suppressed by the following notation, $\bullet$. The bulk-bulk propagators will remain as it is. In the new notations, the diagrams in figure \ref{WittenEX1} are denoted by 
\begin{figure}[H]
\centering
\begin{tikzpicture}
    \begin{scope}[shift={(+2, 0)}]
    \draw (-1, 0) -- (1, 0);
    \node at (-1, 0) {\textbullet};
    \node at (-1, -0.25) {$x_1$};
    \node at (1, 0) {\textbullet};

    \node at (1, -0.25) {$x_2$};

    \node at (0, 0.25) {$y$};

\end{scope}

\begin{scope}[shift={(-2, 0)}]
\node at (0, 0) {\textbullet};
    \node at (0, -0.25) {$x$};
\end{scope}
\end{tikzpicture}
\caption{These are truncations of the diagrams in figure \ref{WittenEX1}. The sum momenta of the bulk-boundary propagators ending on a vertex is denoted by $x_i$ and the mod of the momenta of the intermediate particles is denoted by $y_i$. Therefore for the diagram on the left, we have $x = k_1 + k_2 + k_3$. For the diagram on the right, we have $x_1 = k_1+ k_2$, $x_2 =  k_3 + k_4$, $y = |\bm k_1 + \bm k_2|$.}
\label{WittenEX2}
\end{figure}
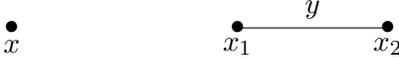

\subsection{Example of a Witten Diagram at one-loop}
In this subsection and the next, we shall briefly review the recursion relations described in \cite{Arkani-Hamed:2017fdk} and construct the loop integrands. We first review the conventional way of performing this computation and describe the recursion relations after that. Similar to the tree level computation, we first decompose the bulk-bulk Green's function as given in \eqref{bulkbulkdS} and then attach the appropriate bulk-boundary propagators (equation \eqref{bulk-bndy}), followed by an integration over the bulk points. Finally, we perform the loop integrals. This gives a clear sense of the order of integration.

The Green's function outlined in equation \eqref{bulkbulkdS} includes terms that consist of $\Theta(\eta_1 - \eta_2)$, $\Theta(\eta_2 - \eta_1)$, and a term that lacks a $\Theta$-function. These can be depicted through the following diagrams\footnote{Note that these diagrams are useful to denote the product of $\Theta-$functions and will not be needed after this subsection. }.
\begin{eqn}\label{notation}
\begin{tikzpicture}[baseline]
\draw[fermion, red] (-1, 0.15) -- (1,0.15);
\node at (-1, 0.15) {$\bullet$};
\node at (1, 0.15) {$\bullet$};
\node at (-1, -0.2) {$\eta_1$};
\node at (1, -0.2) {$\eta_2$};
\end{tikzpicture}
&= \frac{1}{2 y}  \Theta(\eta_1 - \eta_2) e^{- i y (\eta_1 - \eta_2 )}, \\
\begin{tikzpicture}[baseline]
\draw[fermionbar, red] (-1, 0.15) -- (1,0.15);
\node at (-1, 0.15) {$\bullet$};
\node at (1, 0.15) {$\bullet$};
\node at (-1, -0.2) {$\eta_1$};
\node at (1, -0.2) {$\eta_2$};
\end{tikzpicture}
&= \frac{1}{2 y} \Theta(\eta_2 - \eta_1) e^{- i y (\eta_2 - \eta_1 )}, \\
\begin{tikzpicture}[baseline]
\draw[dashed, red] (-1, 0.15) -- (1,0.15);
\node at (-1, 0.15) {$\bullet$};
\node at (1, 0.15) {$\bullet$};
\node at (-1, -0.2) {$\eta_1$};
\node at (1, -0.2) {$\eta_2$};
\end{tikzpicture}
&=\frac{1}{2 y} e^{ i y (\eta_1 + \eta_2 )}~.
\end{eqn}
We now focus on the wave function coefficient at one-loop\footnote{For interactions that are conformally coupled this diagram can correspond to a two-point function in the $\phi^3$ theory in $D = 6$ or a four-point function in $\phi^4$ theory in $D = 4$.}, which can be written as 
\begin{eqn}
\begin{tikzpicture}[baseline]
\draw (-1.5,-0) circle (0.5);
\node at (-2, -0) {\textbullet};
\node at (-1, -0) {\textbullet};
\node at (-2.25, -0) {${\scriptsize{x_1}}$};
\node at (-0.65, -0) {${\scriptsize{x_2}}$};
\node at (-1.5, 0.7) {${\scriptsize{y_1}}$};
\node at (-1.5, -0.7) {${\scriptsize{y_2}}$};
\end{tikzpicture}
=  \int d^3 l_1  \intnsinf d\eta_1 d\eta_2\, e^{i \eta_1 x_1} e^{i \eta_2 x_2} G(\eta_1, \eta_2, y_1) G(\eta_1, \eta_2, y_2)~.
\end{eqn}
This diagram is often known as the \textit{bubble} diagram. From the equation above, we have to evaluate a product of Green's functions, containing the $\Theta$ functions. Upon multiplying two such terms, ones with the opposite directions of $\Theta$ no longer survive as $\Theta(\eta_1 - \eta_2) \Theta(\eta_2 - \eta_1) = 0$\footnote{ More generally this identity is given as $\Theta(\eta_1 - \eta_2) \eta(\eta_2 - \eta_1) = a \delta(\eta_1 - \eta_2)$~, where $a$ is dependent on the choice of the regularization. }. Hence the only allowed contractions in the notations of \eqref{notation} are given below,
\begin{eqn}
\begin{tikzpicture}

\begin{scope}[shift={(-7.5, 0)}]
\draw (-1.5,-0.2) circle (0.5);
\node at (-2, -0.2) {\textbullet};
\node at (-1, -0.2) {\textbullet};
\node at (0,-0.2) {$=$};
\node at (-2.25, -0.2) {${\scriptsize{x_1}}$};
\node at (-0.65, -0.2) {${\scriptsize{x_2}}$};
\node at (-1.5, 0.55) {${\scriptsize{y_1}}$};
\node at (-1.5, -0.95) {${\scriptsize{y_2}}$};

\draw [decorate,decoration={brace,amplitude=10pt},xshift=3,yshift=-4]
(2.5,-2) -- (2.5,2) node [black,midway,xshift=-0.6cm] 
{};
\node at (1, -0.2) {$\largeeq{\int d^3 l_1}$};
\end{scope}

\begin{scope}[shift={(-4.2, 1)}]
 \draw[fermion, red] (1, 0) arc (0:180:0.5);
 \draw[fermion, red] (1, 0) arc (0:-180:0.5);
 \node at (1.5,0) {$-$};
 \node at (1, 0) {\textbullet};
\node at (0, 0) {\textbullet};
\end{scope}

\begin{scope}[shift={(-2.2, 1)}]
 \draw[fermion, red] (1, 0) arc (0:180:0.5);
 \draw[dashed, red] (1, 0) arc (0:-180:0.5);
 \node at (1.5,0) {$-$};
 \node at (1, 0) {\textbullet};
\node at (0, 0) {\textbullet};
\end{scope}

\begin{scope}[shift={(-0.2, 1)}]
 \draw[dashed, red] (1, 0) arc (0:180:0.5);
 \draw[fermion, red] (1, 0) arc (0:-180:0.5);
 \node at (1, 0) {\textbullet};
\node at (0, 0) {\textbullet};
\end{scope}

\begin{scope}[shift={(-4, -1)}]
\node at (-.5, 0) {$+$};
 \draw[fermionbar, red] (1, 0) arc (0:180:0.5);
 \draw[fermionbar, red] (1, 0) arc (0:-180:0.5);
 \node at (1.5,0) {$-$};
 \node at (1, 0) {\textbullet};
\node at (0, 0) {\textbullet};
\end{scope}

\begin{scope}[shift={(-2, -1)}]
 \draw[fermionbar, red] (1, 0) arc (0:180:0.5);
 \draw[dashed, red] (1, 0) arc (0:-180:0.5);
 \node at (1.5,0) {$-$};
 \node at (1, 0) {\textbullet};
\node at (0, 0) {\textbullet};
\end{scope}

\begin{scope}[shift={(0, -1)}]
 \draw[dashed, red] (1, 0) arc (0:180:0.5);
 \draw[fermionbar, red] (1, 0) arc (0:-180:0.5);
 \node at (1, 0) {\textbullet};
\node at (0, 0) {\textbullet};
\end{scope}

\begin{scope}[shift={(2, 0)}]
\node at (-0.5,0) {$+$};
 \draw[dashed, red] (1, 0) arc (0:180:0.5);
 \draw[dashed, red] (1, 0) arc (0:-180:0.5);
 \node at (1, 0) {\textbullet};
\node at (0, 0) {\textbullet};
\draw [decorate,decoration={brace,amplitude=10pt, mirror},xshift=3,yshift=-4]
(1.5,-2) -- (1.5,2) node [black,midway,xshift=+0.6cm] 
{};
\end{scope}
\end{tikzpicture}~.
\label{oneloop}
\end{eqn}
The integrals over $\eta$ are of two kinds, one which has a $\Theta$ function,
\be
\intnsinf d\eta_1 d\eta_2\, e^{i a \eta_1} e^{i b \eta_2} \Theta(\eta_1 - \eta_2) = \intnsinf d\eta_1 e^{i a \eta_1} \int_{-\infty}^{\eta_1} d\eta_2 e^{i b \eta_2}= -\frac{1}{(a+b)b}  ~,
\ee
and one which does not, 
\be
\intnsinf d\eta_1 d\eta_2 e^{i a \eta_1} e^{i b \eta_2}  = - \frac{1}{a b}~.
\ee
The terms generated by these diagrams have spurious poles which cancel amongst each other after summing up all the terms. For example, the contribution from the diagram $\begin{tikzpicture}[baseline]
 \draw[dashed, red] (1, 0) arc (0:180:0.5);
 \draw[fermionbar, red] (1, 0) arc (0:-180:0.5);
 \node at (1, 0) {\textbullet};
\node at (0, 0) {\textbullet};
\end{tikzpicture}$ evaluates to the following  expression 
\begin{eqn}
\begin{tikzpicture}[baseline]
 \draw[dashed, red] (1, 0) arc (0:180:0.5);
 \draw[fermionbar, red] (1, 0) arc (0:-180:0.5);
 \node at (1, 0) {\textbullet};
\node at (0, 0) {\textbullet};
\end{tikzpicture}
= \frac{1}{4 y_1 y_2(x_1 + x_2 + 2 y_1)(x_2 +  y_1 + y_2)}~,
\end{eqn}
which contain the poles $\frac{1}{y_1} \frac{1}{y_2}$~. However, after summing up all the diagrams in \eqref{oneloop} we see that these poles disappear implying that they are spurious. The final integrand for the bubble diagram is given as,
\be\label{oneloop-1-simple}
\begin{tikzpicture}[baseline]
\draw (-1.5,-0) circle (0.5);
\node at (-2, -0) {\textbullet};
\node at (-1, -0) {\textbullet};
\node at (-2.25, -0) {${\scriptsize{x_1}}$};
\node at (-0.65, -0) {${\scriptsize{x_2}}$};
\node at (-1.5, 0.7) {${\scriptsize{y_1}}$};
\node at (-1.5, -0.7) {${\scriptsize{y_2}}$};

\end{tikzpicture}
=  \int d^3 l_1 \frac{1}{(x_1 + x_2)(x_1 + y_1 + y_2)(x_2 + y_1 + y_2)} \Bigg[ \frac{1}{x_1 + x_2  + 2y_1} + \frac{1}{x_1 + x_2  + 2y_2}  \Bigg]~.
\ee
Since we are interested in computing the loop integrals above and also have to regularize them, we must use an expansion that avoids integrating terms with spurious poles. Such an expansion can be derived using {\it old fashioned perturbation theory} as advocated in \cite{Arkani-Hamed:2017fdk} and reviewed in the following section. 

\subsection{Recursion Relations for Scalars} \label{sec:recursion}
We now review the recursion relation developed in \cite{Arkani-Hamed:2017fdk} which shall be used to construct some of the loop integrands. This recursion relation uses the one-point graph 
\bes
\begin{tikzpicture}[baseline]
    \node at (0, 0) {\textbullet}; 
    \node at (0, 0.25) {$x$}; 
\end{tikzpicture} = \frac{1}{x}~,
\ees
as a seed and generates the whole diagram by attaching edges and other vertices. We first note that any Witten diagram $\Psi$ of interest can be reduced to the following equation (after stripping off the momentum-conserving delta function)
\be\label{nimabasis}
\Psi = \intnsinf \prod_v d\eta_v e^{i x_v \eta_v} \prod_e G(\eta_e, \eta_e', y_e)~.
\ee
 Since the boundary ($\eta \to 0$, $\eta \to -\infty$) value of the Green's function is zero, the boundary value of the whole integrand in equation \eqref{nimabasis} is zero. This allows us to simplify the integral by using integration by parts,
\be\label{trick1}
0 = \intnsinf  \prod_v d\eta_v\  \hat \Delta \Big[ e^{\sum_v i x_v \eta_v} \prod_e G(\eta_e, \eta'_e, y_e)\Big]~,
\ee
where $\hat \Delta = - i \sum_{v} \p_{\eta_v}$~. The differential operator, $\hat\Delta$, is a linear operator and it will act on both parts of the integrand, namely, the bulk-boundary propagators and also the bulk-bulk propagators. Upon acting on the bulk-boundary propagators, it gives us the total energy, i.e.,
\be\label{trick2}
 \intnsinf  \prod_v d\eta_v \hat \Delta \Big[ e^{\sum_v i x_v \eta_v} \Big] \prod_e G(y_e, \eta_e, \eta'_e) = \Psi \sum_v x_v ~.
\ee
Its action on the bulk-bulk propagators is more interesting. For that we first study its action on a particular propagator, say $G(\eta, \eta', y)$ as given in \eqref{bulkbulkdS}. Since the differential operator is a function of $\eta+ \eta'$, it will annihilate the part of $G(\eta, \eta', y)$ that contains $\eta - \eta'$. Therefore, the only term that will remain is given as 
\be\label{Deltasimp}
\hat \Delta G(\eta, \eta', y) = - e^{i y (\eta + \eta')}~,
\ee
which simply modifies the bulk-boundary propagators that were already present! Hence the action of $\hat \Delta$ effectively clips the bulk-bulk propagator and adds a factor of $y$ to the vertices at its edge. Thus by using integration by parts, we can obtain the value of the integral without having to explicitly integrate the $\Theta$ functions. For this trick to work it is imperative for the bulk-bulk propagators to satisfy Cauchy's exponential functional equation. Let us study a simple example of this algorithm. Consider the diagram corresponding to the four-point function, denoted by $\Psi_4^{(0)}$,
\be
\Psi_4^{(0)}(x_1, y, x_2) =
\begin{tikzpicture}[baseline]
    \draw (-1, 0) -- (1, 0);
    \node at (-1, 0) {\textbullet};
    \node at (1, 0) {\textbullet};
    \node at (-1, -0.25) {$x_1$};
    \node at (1, -0.25) {$x_2$};
    \node at (0, 0.25) {$y$};
\end{tikzpicture}~.
\ee
By following the procedure described above, we have
\be
(x_1 + x_2) \Psi_4^{(0)} = 
\begin{tikzpicture}[baseline]
    
    \node at (-1, 0) {\textbullet};
    \node at (1, 0) {\textbullet};
    \node at (-1, -0.25) {$x_1 + y$};
    \node at (1, -0.25) {$x_2 + y$};
\end{tikzpicture}
= \frac{1}{(x_1 + y)(x_2 + y)} ~.
\ee
This implies that the value of the diagram is given as
\be
\Psi_4^{(0)}(x_1, y, x_2) = \frac{1}{(x_1 + x_2) (x_1 + y)(x_2 + y)}
\ee
To understand the algorithm better, let us consider the five-point function,
\begin{eqn}
    \begin{tikzpicture}[baseline]
   \draw[thick, black] (-3, 1) -- (3, 1); 
 \draw[fermion, black] (-2, 1) -- (-1.5, -1.5);
 \draw[fermion, black] (2, 1) -- (1.5, -1.5);
 \draw[fermion, black] (-1, 1) -- (-1.5, -1.5);
 \draw[fermion, black] (1, 1) -- (1.5, -1.5);
 \draw[fermion, black] (0, 1) -- (0, -1.5);
 \node at (-2.1, 0) {$\bm k_1$};
 \node at (-0.8, 0) {$\bm k_2$};
\node at (0.8, 0) {$\bm k_3$};
\node at (2.1, 0) {$\bm k_4$};
\node at (0.25, -0.5) {$\bm k_5$};
\draw[fermion, black] (-1.5, -1.5) -- (1.5, -1.5);
\node at (-1.5, -1.75) {$\eta_1$};
\node at (1.5, -1.75) {$\eta_2$};
\node at (0, -1.75) {$\eta_3$};
    \end{tikzpicture}
\end{eqn}
and denote it by $L(x_1, y_1, x_2, y_2, x_3)$. Here $x_1 = k_1 + k_2,\ x_2 = k_5,\ x_3 = k_3 + k_4$ and $y_1 = |\bm k_1 + \bm k_2|$ and $y_2 = |\bm k_3 + \bm k_1 + \bm k_2|$~. In the condensed notation, this diagram is
\begin{eqn}\label{L-5point}
 L(x_1, y_1, x_2, y_2, x_3) 
 \equiv \begin{tikzpicture}[baseline]
    \draw (-2, 0) -- (2, 0);
    \node at (-2, 0) {\textbullet};
    \node at (-2, -0.25) {$x_1$};
    \node at (-1, +0.25) {$y_1$};
    \node at (0, 0) {\textbullet};
    \node at (0, -0.25) {$x_2$};
    \node at (1, +0.25) {$y_2$};
    \node at (2, 0) {\textbullet};
    \node at (2, -0.25) {$x_3$};
   \end{tikzpicture}~.
\end{eqn}
By following the procedure above we obtain the following recursion relation 
\begin{eqn}
(x_1 + x_2 + x_3) L &= 
\begin{tikzpicture}[baseline]
    \draw (0, 0) -- (2, 0);
    \node at (-2, 0) {\textbullet};
    \node at (-2, -0.25) {$x_1 + y_1$};
    
    \node at (0, 0) {\textbullet};
    \node at (0, -0.25) {$x_2 + y_1$};
    \node at (1, +0.25) {$y_2$};
    \node at (2, 0) {\textbullet};
    \node at (2, -0.25) {$x_3$};
   \end{tikzpicture}
   + 
   \begin{tikzpicture}[baseline]
    \draw (-2, 0) -- (0, 0);
    \node at (-2, 0) {\textbullet};
    \node at (-2, -0.25) {$x_1$};
    
    \node at (0, 0) {\textbullet};
    \node at (0, -0.25) {$x_2 + y_2$};
    \node at (-1, +0.25) {$y_1$};
    \node at (2, 0) {\textbullet};
    \node at (2, -0.25) {$x_3+y_2$};
   \end{tikzpicture} ~,
\end{eqn}
where two of the diagrams are evaluated in the previous example and therefore the value of $L(x_1, y_1, x_2, y_2, x_3)$ becomes
\be
 L = \frac{1}{x_1 + x_2 + x_2} \left[ \frac{1}{x_1 + y_1} \Psi_4^{(0)}(x_2+y_1, y_2, x_3) + \Psi_4^{(0)}(x_1, y_1, x_2 + y_2) \frac{1}{x_3 + y_2}  \right]~.
\ee

 We can also demonstrate the use of the recursion relations in the case of the one-loop diagram discussed before \eqref{oneloop}, 
\be
\begin{tikzpicture}[baseline]
\draw (-1.5,-0) circle (0.5);
\node at (-2, -0) {\textbullet};
\node at (-1, -0) {\textbullet};
\node at (-2.25, -0) {${\scriptsize{x_1}}$};
\node at (-0.65, -0) {${\scriptsize{x_2}}$};
\node at (-1.5, 0.7) {${\scriptsize{y_1}}$};
\node at (-1.5, -0.7) {${\scriptsize{y_2}}$};
\end{tikzpicture}
= \frac{1}{x_1 + x_2} \Big[ \Psi_4^{(0)}(x_1 + y_1, y_1, x_2 + y_1) + \Psi_4^{(0)}(x_1 + y_2, y_2, x_2 + y_2)  \Big]~.
\ee
This matches with the expression \eqref{oneloop-1-simple} which was evaluated by explicitly summing all the terms.

We pause to mention an important physical point about the poles in the expression of the wave function. Notice that by using this recursion relation we always end up with the denominator containing the total energy ($E_{tot}$) entering the diagram. For example, in the case of the one-loop diagram in \eqref{oneloop-1-simple} we have the pole $\frac{1}{x_1 + x_2}$. This factor of $\frac{1}{E_{tot}}$ is a universal factor and is present in every diagram which is easy to see by using the recursion relation in any general graph \cite{Arkani-Hamed:2017fdk}\footnote{The other poles are also of physical significance and we refer the reader to \cite{Arkani-Hamed:2017fdk} for a discussion on this.}. By analytically continuing the energy and taking the residue at this pole we obtain the corresponding S-matrix for high energy scattering in flat space and hence this is known as the flat space limit \cite{Raju:2012zr}\footnote{For tree level diagrams the residue at this pole exactly corresponds to the flat space S-matrix. For loop-level diagrams, the flat space limit gives the corresponding loop-level integrand of the S-matrix but with the integration over $l^0$ already performed.}. This is because this pole effectively restores energy conservation, which was broken because of the boundary at $\eta \to 0$. In other words, the scattering process ``stops feeling the effect of the boundary'' in this limit. It is important to note that there is no pole in the physical space of the momenta and for seeing these features the momenta have to be analytically continued. We encourage the interested reader to refer to \cite{Arkani-Hamed:2017fdk, Benincasa:2022gtd} for a discussion on this issue. 

\section{Bulk Witten Diagrams}
In this paper, we go beyond the work in \cite{Arkani-Hamed:2017fdk, Albayrak:2020isk}  and evaluate the integrals for certain two-loop diagrams and also for the one-loop triangle diagram. These will be relevant for the case of $\phi^4$ theory in $D=4$ and for $\phi^3$ theory in $D=6$ respectively. We shall use the recursion relations \cite{Arkani-Hamed:2017fdk} reviewed in the previous section to write the loop integrands.

\subsection{$\phi^4$ in dS$_4$}
The diagrams that contribute to the two-point function in  first and second order in perturbation theory are given below,
\begin{figure}[H]
\centering
 \begin{tikzpicture}

\begin{scope}[shift={(-3.5, 0)}]
\draw (0,0.5) circle (0.5);
\node at (0, 0) {\textbullet};
\node at (0, -0.25) {$x_1$};
\draw[dashed, thick] (-1, -0) -- (1, -0);
\end{scope}

\begin{scope}[shift={(-1, 0)}]
\draw (0,0.5) circle (0.5);
\draw (0,1.5) circle (0.5);
\node at (0, 0) {\textbullet};
\node at (0, -0.25) {$x_1$};
\draw[dashed, thick] (-1, -0) -- (1, -0);
\node at (0, 1) {\textbullet};
\node at (0, 0.75) {$x_2$};
\end{scope}

\begin{scope}[shift={(2.5, 0)}]
\draw (-1,0.5) circle (0.5);
\draw (1,0.5) circle (0.5);
\node at (-1, 0) {\textbullet};
\node at (-1, -0.25) {$x_1$};
\node at (1, 0) {\textbullet};
\node at (1, -0.25) {$x_2$};
\node at (0, -0.25) {$y$};
\draw[dashed, thick] (-2, -0) -- (-1, -0);
\draw[dashed, thick] (2, -0) -- (1, -0);
\draw (-1, -0) -- (+1, -0);
\end{scope}

\begin{scope}[shift={(7, 0)}]
\draw (0,0) circle (1);
\draw (-1, 0) -- (1, 0);
\draw[dashed, thick] (-2, 0) -- (-1, 0);
\draw[dashed, thick] (1, 0) -- (2, 0);
\node at (-1, -0) {\textbullet};
\node at (1, -0) {\textbullet};
\node at (-1.25, -0.25) {${\scriptsize{x_1}}$};
\node at (1.25, -0.25) {${\scriptsize{x_2}}$};
\node at (0, 1.25) {${\scriptsize{y_1}}$};
\node at (0, 0.25) {${\scriptsize{y_2}}$};
\node at (0, -1.25) {${\scriptsize{y_3}}$};
\end{scope}

\end{tikzpicture}

\caption{ The bulk-boundary propagators are denoted by a bold dashed line in order to avoid drawing the boundary of spacetime. 
}
\end{figure}
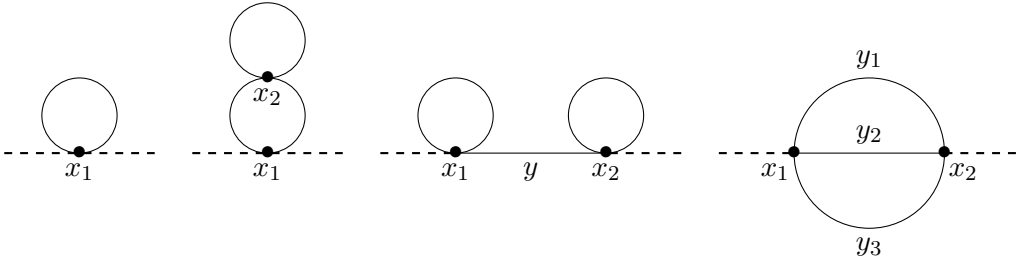

\subsubsection{Cactus Diagrams}
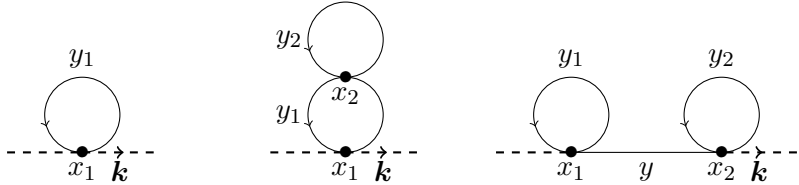
\begin{figure}[H]
\centering
    \begin{tikzpicture}
    \begin{scope}[shift={(-3.5, 0)}]
\draw[ fermion, black] (0,0.5) circle (0.5);
\node at (0, 0) {\textbullet};
\node at (0, -0.25) {$x_1$};
\draw[dashed, thick] (-1, -0) -- (0, -0);
\draw[thick, dashed, fermion, black] (0, -0) -- (1, -0);
\node at (0.5, -0.25) {$\bm k$};
\node at (0, 1.25) {$y_1$};
\end{scope}

\begin{scope}[shift={(0, 0)}]
\draw[fermion, black] (0,0.5) circle (0.5);
\draw[fermion, black] (0,1.5) circle (0.5);
\node at (0, 0) {\textbullet};
\node at (0, -0.25) {$x_1$};
\draw[dashed, thick] (-1, -0) -- (0, -0);
\draw[thick, fermion, dashed, black] (0, -0) -- (1, -0);
\node at (0.5, -0.25) {$\bm k$};
\node at (0, 1) {\textbullet};
\node at (0, 0.75) {$x_2$};
\node at (-0.75, 0.5) {$y_1$};
\node at (-0.75, 1.5) {$y_2$};
\end{scope}

\begin{scope}[shift={(4, 0)}]
\draw[fermion, black] (-1,0.5) circle (0.5);
\draw[fermion, black] (1,0.5) circle (0.5);
\node at (-1, 0) {\textbullet};
\node at (-1, -0.25) {$x_1$};
\node at (1, 0) {\textbullet};
\node at (1, -0.25) {$x_2$};
\node at (0, -0.25) {$y$};
\draw[dashed, thick] (-2, -0) -- (-1, -0);
\draw[dashed, thick, fermionbar, black] (2, -0) -- (1, -0);
\node at (1.5, -0.25) {$\bm k$};
\draw (-1, -0) -- (+1, -0);
\node at (-1, 1.25) {$y_1$};
\node at (1, 1.25) {$y_2$};
\end{scope}
\end{tikzpicture}
\caption{Contributions from the Cactus diagrams at one-loop and two-loop. }
\label{2cactus}
\end{figure}
The recursion relations in the previous section can also be used to write the loop integrand for the cactus diagram. However, it requires one to carefully regularize some integrals as shown below. Using the diagrammatic rules, the expression for the one-loop diagram above is given as,
\be\label{cactus2integrand}
R_{(1)} = \intnsinf d\eta\, e^{ i x_1 \eta} G(\eta, \eta, y_1)~,
\ee
with $x_1 = 2 |\bm k|$ and $y_1 = |\bm l|$. Following the method described in equation \eqref{trick1} we can consider the integral of the total derivative (in $\eta$) of the integrand in equation \eqref{cactus2integrand}. However,the Green's function at coincident points $G(\eta, \eta, y)$ is not zero as $\eta \to -\infty$\footnote{\bes
G(\eta, \eta, y) = (1 - e^{ 2 i y \eta})/(2y)~.
\ees } and  therefore we need to regularize it using the point splitting procedure and then taking the limit $\lim_{\eta' \to \eta}G(\eta, \eta', y)$, which gives $0$ as $\eta \to -\infty$ (following from the boundary value of the Green's function). Thus consider the following regularized expression\footnote{The $\delta(\eta - \eta')$ function in this step should be thought of a distribution.}
\begin{eqn}\label{regularization1}
0 &= (-i)\intnsinf d\eta d\eta' \Big(\frac{\partial}{\partial \eta} + \frac{\partial}{\partial \eta'} \Big) \Big[  e^{i  x_1 \eta}\delta(\eta - \eta') G(\eta, \eta', y_1) \Big]~, \\ 
&=  \intnsinf d\eta d\eta' e^{i x_1 \eta} \delta(\eta - \eta')\left[ x_1 + \Big(\frac{\partial}{\partial \eta} + \frac{\partial}{\partial \eta'} \Big)  G(\eta, \eta', y_1) \right]~.
\end{eqn}
By using the formula \eqref{Deltasimp} and integrating over $\delta(\eta - \eta')$ we obtain
\be\label{cactus1integrand}
 \intnsinf d\eta\, e^{i x_1 \eta} \left[ x_1  - e^{2 i \eta y_1} \right] = 0 \implies R_{(1)} = \frac{1}{x_1 (x_1 + 2y_1)}~.
\ee 
Restoring the loop integral we get the following expression 
\begin{eqn}
    \bm R_{(1)} = \int d^3 l\, \frac{1}{4 k (k + l)}~.
\end{eqn}
Since the integral is spherically symmetric we can easily evaluate it by going to spherical polar coordinates
\be
\bm R_{(1)} = \frac{4\pi}{4 k}  \int_0^\Lambda \frac{l^2 dl}{k + l} ~,
\ee
which gives 
\be\label{tadpole1}\boxed{
\bm R_{(1)} = \frac{\Lambda^2}{4k} - \frac{\Lambda}{2} - \frac{k}{2} \log \frac{k}{\Lambda}~.
}\ee
Notice that the flat space limit only captures the UV part of the result and the finite part of this diagram has a different behavior from the corresponding Feynman diagram in flat space (as this is dependent on the external momentum $\bm k$) unless the theory contains derivative interactions. Divergences appearing as polynomials in a $\Lambda$ are absent in another well-known regularization scheme known as dimensional regularization.

It is interesting to note that this integral can also be evaluated using the Feynman rules by re-writing the bulk-bulk propagator in terms of its ``spectral representation'', 
\be\label{spectral}
R_{(1)} = \frac{2}{\pi} \intnsinf d\eta\, e^{2i k \eta} \intsinf dp \frac{\sin^2(p \eta)}{p^2 - l^2}~,
\ee
and performing the $\eta$-integral, we obtain
\be
R_{(1)} = \frac{1}{2\pi k} \intsinf \frac{dp}{p^2 - l^2} \frac{p^2}{p^2 + k^2} = \frac{1}{4k (k+l)}~,
\ee
which reproduces the same integrand as equation \eqref{cactus1integrand} evaluated using the algorithm. 

The 1PI two-loop cactus diagram (see figure \ref{2cactus}) is more complicated and requires careful regularization for the bulk point integrals as described in appendix \ref{app:cactus2}. Using the result derived in the appendix \eqref{cactusloopintegrand}  the loop integral becomes,
\be
\bm R_{(2)} =\int  \frac{2d^3 l_1 d^3 l_2}{x_1(x_1 + 2 y_1)} \Big[ \frac{1}{( 2 y_1)(2y_2 + 2 y_1)(x_1 + 2 y_1)}
+ \frac{2}{ (x_1 + 2 y_1 + 2 y_2)(x_1 + 2y_1)( 2y_1 + 2 y_2) }  \Big] ~,
\ee
with $x_1 = 2|\bm k|$, $y_1 = |\bm l_1|$ and $y_2 = |\bm l_2|$~. Since the integrals are spherically symmetric, it is straightforward to perform them in the same way as that of the single loop. Also, the $\bm l_1$ and $\bm l_2$ integrations are independent of each other and therefore, the integration measure simply becomes $d^3 l_1 d^3 l_2 = (4\pi)^2 l_1^2 l_2^2 dl_1 dl_2$. The order of integration does not matter but we find that it is convenient to perform the integration over the outermost loop $(\bm l_2)$ and then go inwards ($\bm l_1$). 

Upon performing the integrals, we obtain the following result for $\bm R_{(2)}$,
\be\label{tadpole21}\boxed{
\bm R_{(2)} = \frac{\pi ^2 \Lambda ^2}{4 k} \Big[3 -2 \log \frac{k}{\Lambda}\Big]
+  \pi^2 \Lambda  \Big[5+2 \log \frac{k}{\Lambda}\Big]
-\frac{90\pi^2 k}{72}  \Big[\frac{\pi^2}{3}+ \frac{13}{10} + \Bigl( \log \frac{k}{\Lambda} +\frac{1}{3}  \Bigr)^2 \Big]
}\ee

Apart from the conventional 1PI diagrams we also have another diagram at 2-loop as shown in figure \ref{2cactus}. Unlike the usual Feynman diagrams in QFT, it is not obvious that the connected Witten diagrams can be trivially obtained from the 1PI Witten diagrams as one has to integrate over the bulk points that contain the bulk-bulk propagators. This utilizes the procedure we described for the diagrams above and we obtain the following relation
\begin{eqn}
\begin{tikzpicture}[baseline]
\draw[black] (-1,0.5) circle (0.5);
\draw[black] (1,0.5) circle (0.5);
\node at (-1, 0) {\textbullet};
\node at (-1, -0.25) {$x_1$};
\node at (1, 0) {\textbullet};
\node at (1, -0.25) {$x_2$};
\node at (0, -0.25) {$y$};
\draw (-1, -0) -- (+1, -0);
\node at (-1, 1.25) {$y_1$};
\node at (1, 1.25) {$y_2$};
\end{tikzpicture}
&=\frac{1}{(x_1 + x_2)} \Big[ \begin{tikzpicture}[baseline]
\draw[ black] (1,0.5) circle (0.5);
\node at (-1, 0) {\textbullet};
\node at (-1, -0.25) {$x_1+2y_1$};
\node at (1, 0) {\textbullet};
\node at (1, -0.25) {$x_2$};
\draw (-1, 0) -- (1, 0);
\node at (0, 0.25) {$y$};
\end{tikzpicture} 
+ \begin{tikzpicture}[baseline]
\draw[ black] (-1,0.5) circle (0.5);
\node at (-1, 0) {\textbullet};
\node at (-1, -0.25) {$x_1$};
\node at (1, 0) {\textbullet};
\node at (1, -0.25) {$x_2+2y_2$};
\draw (-1, 0) -- (1, 0);
\node at (0, 0.25) {$y$};
\end{tikzpicture}  \\
&\qquad\qquad 
+ \begin{tikzpicture}[baseline]
\draw[black] (-1,0.5) circle (0.5);
\draw[black] (1,0.5) circle (0.5);
\node at (-1, 0) {\textbullet};
\node at (-1, -0.25) {\scriptsize{$x_1+y+2y_1$}};
\node at (1, 0) {\textbullet};
\node at (1, -0.25) {\scriptsize{$x_2+y+2y_2$}};
\end{tikzpicture} \Big] ~.
\end{eqn}
It should be noted that only the last diagram splits into 2-1PI diagrams. Using the algorithm we can write the final loop integrand,
\begin{eqn}
&\int d^3 l_1 d^3 l_2\, R_{(1, 1)}\\
&= \frac{1}{(x_1 + x_2)(x_1 + y + 2y_1)(x_2 + y + 2y_2)} \\
&\times\Big[ \frac{2}{(x_1 + y)(x_1 + x_2 + 2 y_1)} + \frac{2}{(x_1 + x_2 + 2y_1 + 2 y_2)(x_1 + x_2 + 2y_1)}+ \frac{1}{(x_1 + y)(x_2 + y)} \Big]~,
\end{eqn}
where $x_1 = k$, $x_2 = k$ and $y = k, y_1 = l_1, y_2  = l_2$. Since the diagram does not completely reduce to a product of $\bm R_{(1)}$ it implies that studying the conventional 1PI diagrams are not sufficient. The final loop integral for this diagram is given as
\begin{framed}
\begin{eqn}\label{tadpole22}
    \bm R_{(1, 1)} &= \int d^3 l_1 d^3 l_2 \frac{1}{8k(k+l_1)(k+l_2)} \Big[ \frac{1}{2k(k+l_1)}+ \frac{1}{2(k+ l_1 + l_2)(k+l_1)} + \frac{1}{4k^2}\Big] ~,\\
    &=\frac{\pi ^2 \Lambda ^4}{8 k^3}+\frac{\pi^2\Lambda^2}{4 k} \Big(2 \log\frac{k}{\Lambda} +3\Big)-\frac{\pi ^2 \Lambda }{4} 
    - \frac{1}{2} \pi^2 k \Big(  \log \frac{k}{\Lambda} - \frac{1}{2} \Big)^2~.
\end{eqn}
\end{framed}

Although these diagrams have been regularized for interactions that are conformally invariant, it is possible to use the same regularization technique for any time-dependent interaction as explained in section \ref{sec:Concl}.

\subsubsection{Sunset Diagram}
A more complicated diagram at two-loop is the sunset diagram. The integrand for the sunset diagram can be written in a convenient form using the algorithm described in section \ref{sec:review},
\be
\begin{tikzpicture}[baseline]
\draw (0,0) circle (1);
\draw (-1, 0) -- (1, 0);
\draw[dashed, thick] (-2, 0) -- (-1, 0);
\draw[thick, dashed, fermion, black] (1, -0) -- (2, -0);
\node at (1.5, 0.25) {$\bm k$};
\node at (-1, -0) {\textbullet};
\node at (1, -0) {\textbullet};
\node at (-1.25, -0.25) {${\scriptsize{x_1}}$};
\node at (1.25, -0.25) {${\scriptsize{x_2}}$};
\node at (0, 1.25) {${\scriptsize{y_1}}$};
\node at (0, 0.25) {${\scriptsize{y_2}}$};
\node at (0, -1.25) {${\scriptsize{y_3}}$};
\end{tikzpicture}
\equiv  \int d^3 l_1 d^3 l_2 \sum_{i =1}^{3} \sum_{j = 1}^2 \mathcal Y_{ij}~,
\ee
where the terms are given as
\begin{eqn}
\mathcal Y_{11} &= \frac{1}{E_T} \frac{1}{x_1 + x_2 + 2 y_3} \frac{1}{x_1 + x_2 + 2(y_2 + y_3)} \frac{1}{x_1 + y_1 + y_2 + y_3} \frac{1}{x_2 + y_1 + y_2 + y_3}, \\
\mathcal Y_{12} &= \frac{1}{E_T} \frac{1}{x_1 + x_2 + 2 y_3} \frac{1}{x_1 + x_2 + 2(y_1 + y_3)} \frac{1}{x_1 + y_1 + y_2 + y_3} \frac{1}{x_2 + y_1 + y_2 + y_3} , \\
\mathcal Y_{21} &= \frac{1}{E_T} \frac{1}{x_1 + x_2 + 2 y_1} \frac{1}{x_1 + x_2 + 2(y_1 + y_2)} \frac{1}{x_1 + y_1 + y_2 + y_3} \frac{1}{x_2 + y_1 + y_2 + y_3} , \\
\mathcal Y_{22} &= \frac{1}{E_T} \frac{1}{x_1 + x_2 + 2 y_1} \frac{1}{x_1 + x_2 + 2(y_1 + y_3)} \frac{1}{x_1 + y_1 + y_2 + y_3} \frac{1}{x_2 + y_1 + y_2 + y_3}, \\
\mathcal Y_{31} &= \frac{1}{E_T} \frac{1}{x_1 + x_2 + 2 y_2} \frac{1}{x_1 + x_2 + 2(y_1 + y_2)} \frac{1}{x_1 + y_1 + y_2 + y_3} \frac{1}{x_2 + y_1 + y_2 + y_3}, \\
\mathcal Y_{32} &= \frac{1}{E_T} \frac{1}{x_1 + x_2 + 2 y_2} \frac{1}{x_1 + x_2 + 2(y_2 + y_3)} \frac{1}{x_1 + y_1 + y_2 + y_3} \frac{1}{x_2 + y_1 + y_2 + y_3}~.
\end{eqn}
We use the notation $E_T = x_1 + x_2~$. The values of $x_i, y_i$ can be written in terms of the external momenta $\bm k$ and the loop momenta $\bm l_1, \bm l_2$, 
    \bes
x_1 = x_2 = |\bm k|, \qquad 
y_1 = |\bm l_1|, \quad 
y_2 = |\bm l_1 + \bm k - \bm l_2|, \quad 
y_3 = |\bm l_2|~.
\ees
We now proceed onto evaluating this integral. The value of integrals over $\mathcal Y_{ij}$ shall be denoted by $\mathcal T_{ij}$, i.e., $\mathcal T_{ij} = \int d^3 l_1 d^3 l_2 \,\mathcal Y_{ij}$~. By a simple change of variables, we see that the number of independent integrals are just three, which are given as, $\mathcal T_{11}, \mathcal T_{12}$ and $\mathcal T_{31}$. This can be seen by making the following substitution $\bm l_1 \leftrightarrow \bm l_2$ and $\bm k \to -\bm k$~. It is also interesting to note that the integrals $\mathcal T_{12}$ and $\mathcal T_{31}$ are related to $\mathcal T_{11}$ by change of variables. To see this, in $\mathcal T_{12}$ we can use the substitution $-\bm l_1' = \bm l_1 + \bm k - \bm l_2$ and in $\mathcal T_{31}$, we use $\bm l_2' = \bm l_1 + \bm k - \bm l_2$. These then give us $\mathcal T_{12} =  \mathcal T_{11}$ and $\mathcal T_{31} = \mathcal T_{12}$. This means the final contribution to the sunset diagram is simply $6 \mathcal T_{11}$. This shows that there is further simplification at the level of the integrals than in the integrands\footnote{This suggests that the loops integrals over the triangulations of the cosmological polytope \cite{Arkani-Hamed:2017fdk} can give rise to simpler geometric structures. It might be interesting to study the loop integrals from this viewpoint.}. We now proceed to the evaluation of $\mathcal T_{11}$~.

\subsubsection*{Evaluation of $\mathcal T_{11}$}
By plugging in the values of $x_i$'s and $y_i$'s in $\mathcal Y_{11}$, we obtain the following 
\be
\mathcal T_{11} = \int d^3 l_1 d^3 l_2 \frac{1}{4E_T} \frac{1}{k + l_2} \frac{1}{k+ |\bm l_1 + \bm k - \bm l_2| + l_2} \frac{1}{(k + l_1 + |\bm l_1 + \bm k - \bm l_2| + l_2)^2}.
\ee
It is easier to perform the integral over $\bm l_1$ first, 
\begin{eqn}
\mathcal T_{11} =  \int d^3 l_2 \frac{1}{4E_T} \frac{1}{k + l_2}  \int d^3 l_1 \frac{1}{k+ |\bm l_1 + \bm k - \bm l_2| + l_2} \frac{1}{(k + l_1 + |\bm l_1 + \bm k - \bm l_2| + l_2)^2}.
\end{eqn}
The advantage of performing the integrals in this order is that the integrals are manifestly axisymmetric and therefore we can align the vector $\bm k - \bm l_2$ (which is to be held constant while integrating over $\bm l_1$) about the $z$-axis of $\bm l_1$. By doing this, the $\bm l_1$ integral becomes 
\begin{eqn}
&\int  \frac{2\pi d l_1 d\cos\q_1}{k+ l_2 + \sqrt{l_1^2 + |\bm k - \bm l_2|^2 + 2 l_1 |\bm k - \bm l| \cos \q_1 }}
\frac{1}{(k + l_1 + l+2 + \sqrt{l_1^2 + |\bm k - \bm l_2|^2 + 2 l_1 |\bm k - \bm l| \cos \q_1 })^2}.
\end{eqn}
This integral can be performed by regulating the $l_1$ integral by a hard cutoff $\Lambda$~.  One can similarly perform the integral over $\bm l_2$ by aligning the vector $\bm k$ with the $z$-axis of $\bm l_2$~. The final result after performing the two integrals is given as 
\begin{eqn}\label{sunset}
\mathcal T_{11} 
 = \frac{\pi k^2}{E_T} \left[\frac{\Lambda^2}{6k^2} \left(6-\pi ^2\right) -\frac{3\Lambda}{2 k} +  \left(\log \frac{k}{\Lambda} -\frac{1}{12}\right)^2+\frac{1}{18} \Big(4 \pi ^2+\frac{95}{8} \Big)\right]~.
\end{eqn}
By substituting $E_T = 2 k$ we obtain the final value of the sunset diagram,
\be\boxed{
\begin{tikzpicture}[baseline]
\draw (0,0) circle (1);
\draw (-1, 0) -- (1, 0);
\draw[dashed, thick] (-2, 0) -- (-1, 0);
\draw[dashed, thick, fermion, black] (1, -0) -- (2, -0);
\node at (1.5, 0.25) {$\bm k$};
\node at (-1, -0) {\textbullet};
\node at (1, -0) {\textbullet};
\node at (-1.25, -0.25) {${\scriptsize{x_1}}$};
\node at (1.25, -0.25) {${\scriptsize{x_2}}$};
\node at (0, 1.25) {${\scriptsize{y_1}}$};
\node at (0, 0.25) {${\scriptsize{y_2}}$};
\node at (0, -1.25) {${\scriptsize{y_3}}$};
\end{tikzpicture}
=  \pi k \left[\frac{\Lambda^2}{2k^2} \left(6-\pi ^2\right) -\frac{9\Lambda}{2k}+ 3 \left(\log \frac{k}{\Lambda} -\frac{1}{12}\right)^2+\frac{1}{6} \Big(4 \pi ^2+\frac{95}{8} \Big)\right]~.
}\ee
 The momentum dependence is consistent with the expectation from the CFT 2-point function. The UV divergence in this diagram is as expected from the corresponding diagram in flat space.

 For computing the full two-point function at two loops, we would also require the diagram obtained from the counterterm which would be obtained by renormalizing the one-loop cactus, 
 
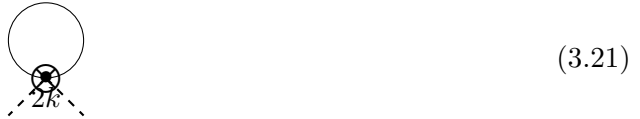
\begin{eqn}
\begin{tikzpicture}[baseline]
\draw (0, 0.5) circle  (0.5);
\node at (0, 0) {\textbullet};
\node at (0, 0) {$\bigotimes$};
\node at (0, -0.25) {$2k$};
\draw[dashed, thick] (0, 0) -- (-0.5, -0.5);
\draw[dashed, thick] (0, 0) -- (0.5, -0.5);
\end{tikzpicture}
 \end{eqn}
We leave the discussion of these diagrams and renormalization schemes in upcoming work.

 The method described above can also be used to study more complicated diagrams of similar kind, such as the four-point function in $\phi^6$ theory in $D =3$,
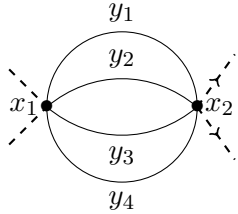
\begin{figure}[H]
\centering
\begin{tikzpicture}[baseline]
\draw (0,0) circle (1);
\draw[dashed, thick] (-1.5, 0.5) -- (-1, 0);
\draw[dashed, thick] (-1.5, -0.5) -- (-1, 0);
\draw[dashed, thick, fermion, black] (1, -0) -- (1.5, 0.7);
\draw[dashed, thick, fermion, black] (1, -0) -- (1.5, -0.7);
\node at (-1, -0) {\textbullet};
\node at (1, -0) {\textbullet};
\node at (-1.3, -0) {${\scriptsize{x_1}}$};
\node at (1.3, -0) {${\scriptsize{x_2}}$};
\node at (0, 1.25) {${\scriptsize{y_1}}$};
\node at (0, 0.65) {${\scriptsize{y_2}}$};
\node at (0, -0.65) {${\scriptsize{y_3}}$};
\node at (0, -1.25) {${\scriptsize{y_4}}$};

\draw (-1, 0) to[out=-40,in=220] (1, 0);
\draw (-1, 0) to[out=40,in=140] (1, 0);
\end{tikzpicture}
\end{figure}

\subsubsection{4-Point Function: Bubble Diagram}
Using a similar technique we can also evaluate the four-point function in $\phi^4$ theory as the leading loop contribution comes from a bubble diagram of the form \cite{Albayrak:2020isk},
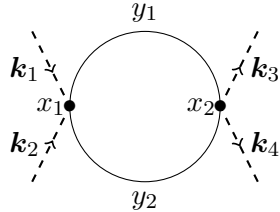
\begin{figure}[H]
\centering
 \begin{tikzpicture}
\draw (0,0) circle (1);
\draw[dashed, fermion, thick, black] (-1.5, 1) -- (-1, 0);
\draw[dashed, fermion, thick, black] (-1.5, -1) -- (-1, 0);
\draw[dashed, fermionbar, thick, black] (1.5, 1) -- (1, 0);
\draw[dashed, fermionbar, thick, black] (1.5, -1) -- (1, 0);
\node at (-1.6, 0.5) {$\bm k_1$};
\node at (-1.6, -0.5) {$\bm k_2$};
\node at (1.6, 0.5) {$\bm k_3$};
\node at (1.6, -0.5) {$\bm k_4$};
\node at (-1, -0) {\textbullet};
\node at (1, -0) {\textbullet};
\node at (-1.25,0) {${\scriptsize{x_1}}$};
\node at (0.75,0) {${\scriptsize{x_2}}$};
\node at (0, 1.25) {${\scriptsize{y_1}}$};
\node at (0, -1.25) {${\scriptsize{y_2}}$};
\end{tikzpicture}
\caption{One-Loop contribution to the Four-Point Function. }
\end{figure}
with $x_1 = k_{12} = |\bm k_1| + |\bm k_2|$,  $x_2 = k_{34} = |\bm k_3| + |\bm k_4|$, $y_1 = |\bm l|$ and $y_2 = |\bm l + \bm k_1 + \bm k_2|$~. These classes of diagrams are commonly known as the \textit{bubble} diagram. The integrand of this follows in a straightforward manner from the algorithm \cite{Arkani-Hamed:2017fdk, Albayrak:2020isk}, 
\begin{eqn}\label{4point-oneloop}
    A_{4}^{(1)} = \frac{1}{(x_1 + x_2)  (x_1 + y_1 + y_2) (x_2 + y_1 + y_2)}\left[ \frac{1}{x_1 + x_2 + 2 y_1} +\frac{1}{x_1 + x_2 + 2 y_2}  \right] ~.
\end{eqn}
The integral is given as 
\begin{eqn}
\bm A_4^{(1)} = \int \frac{d^3l}{E_T  (k_{12} + l + |\bm l + \bm k_1 + \bm k_2|) (k_{34} + l + |\bm l + \bm k_1 + \bm k_2|)} \left[ \frac{1}{E_T + 2 l} +\frac{1}{E_T + 2 |\bm l + \bm k_1 + \bm k_2|}  \right] ~
\end{eqn}
with $E_T = k_{12} + k_{34}$~. There are  two terms appearing in the expression above and by a change of variables $\bm l' = \bm l + \bm k_1 + \bm k_2$, they reduce to the same term 
\be
\bm A_4^{(1)} = \int \frac{2d^3l}{E_T  (k_{12} + l + |\bm l + \bm k_1 + \bm k_2|) (k_{34} + l + |\bm l + \bm k_1 + \bm k_2|)(E_T + 2 l)} ~.
\ee
We can perform this integral in a similar way as before and use 
\bes
|\bm l + \bm k_1 + \bm k_2| = \sqrt{l^2 + (\bm k_1 + \bm k_2)^2 + 2 k |\bm k_1 + \bm k_2| \cos\q}
\ees 
with the measure of the integral given as $d^3l = 2\pi l^2 dl \sin\q d\q$~.
\begin{equation}\label{4ptoneloop}
\boxed{
\begin{aligned}
  \bm A_4^{(1)}  &= \frac{1}{8 k }\Big[\frac{\pi ^2}{3}+\frac{4 k \log 2}{k_{12}+k_{34}}
  +\log ^2\left(\frac{k_{34}-k}{k+k_{12}}\right)+\log ^2\left(\frac{k_{12}-k}{k+k_{34}}\right)-\log ^2\left(\frac{k+k_{12}}{k+k_{34}}\right)
    \\
    &+ 2\text{Li}_2\frac{k+k_{34}}{k-k_{12}}+2 \text{Li}_2\frac{k+k_{12}}{k-k_{34}}
    +\frac{4 k}{k_{12}^2-k_{34}^2} \left(k_{34} \log \frac{k+k_{12}}{\Lambda }-k_{12} \log \frac{k+k_{34}}{\Lambda}\right)
    \Big]
\end{aligned}
}\end{equation}
This result shows that the 4-point function at one-loop (for conformally coupled interactions) has a higher transcendentality (as it contains Li$_2(x)$) than the two-point functions.

\subsection{$\phi^3$ in dS$_6$}
We now turn to the one-loop diagrams in $\phi^3$ theory in $D=6$. The advantage of working in $D=6$ is that it makes the theory conformally coupled and therefore we can use the algorithm described in the previous section to evaluate this.

\subsubsection{One-loop Diagram at 2 points: Bubble Diagram}
The one-loop contribution in the two-point wave function coefficient is obtained from the following diagram\footnote{This diagram was computed in \cite{Albayrak:2020isk} with a wrong integration measure, which is corrected in this paper.}
\begin{figure}[H]
\centering
 \begin{tikzpicture}
\draw (0,0) circle (1);
\draw[dashed, thick] (-2, 0) -- (-1, 0);
\draw[dashed, thick, fermion, black] (1, 0) -- (2, 0);
\node at (1.5, 0.25) {$\bm k$};
\node at (-1, -0) {\textbullet};
\node at (1, -0) {\textbullet};
\node at (-1.25, -0.25) {${\scriptsize{x_1}}$};
\node at (1.25, -0.25) {${\scriptsize{x_2}}$};
\node at (0, 1.25) {${\scriptsize{y_1}}$};
\node at (0, -1.25) {${\scriptsize{y_2}}$};
\end{tikzpicture}
\caption{2-point one-loop graph for $\phi^3$ theory}
\label{fig:2pt-oneloop-phi3}
\end{figure}
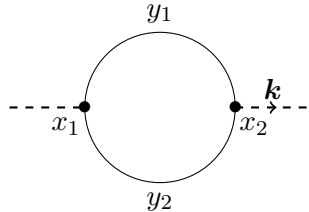
This diagram also falls under the class of the bubble diagrams and therefore has the same integrand as \eqref{4point-oneloop} in the notation of \cite{Arkani-Hamed:2017fdk}, albeit, with different values of $x_1, x_2, y_1$ and $y_2$.
\be
A_{\phi^3, (2)} = \frac{1}{(x_1 + x_2)  (x_1 + y_1 + y_2) (x_2 + y_1 + y_2)}\left[ \frac{1}{x_1 + x_2 + 2 y_1} +\frac{1}{x_1 + x_2 + 2 y_2}  \right] ~,
\ee
with $x_1 = x_2 = |\bm k|$, $y_1 = |\bm l|$ and $y_2 = |\bm l + \bm k|$~. Since the loop integral is performed in $D=6$ and is axisymmetric the measure of the  integral becomes $d^5 l = S_3 l^4 \sin^3 \q d\q dl $, where $S_3$ is the area of the unit 3-sphere. By substituting this and performing the $l$ integral after the $\q$ integral, we obtain the final value of the two-point function at one-loop
\be\label{2pt-phi3-bubble-5dim}\boxed{
\bm A_{\phi^3, (2)} =\frac{\Lambda ^2 S_3}{12 k}
-\frac{\Lambda S_3}{3}
-\frac{k S_3}{120} \Bigl( 7 +50 \log \frac{k}{\Lambda} \Bigr)~.
}\ee
A part of this diagram was also studied using the cosmological cutting rules developed in \cite{Melville:2021lst}. 

\subsubsection{One-loop Diagram at 3 points: Triangle Diagram}\label{sec:triangle}
The contribution to the three-point wave function coefficient at one loop is given by the triangle diagram~.
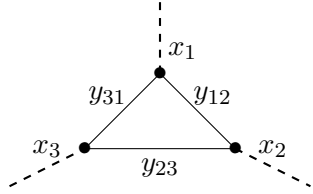
\begin{figure}[H]
 \begin{center}
  \begin{tikzpicture}
   \draw (1, 0) -- (-1, 0);
\draw (1, 0) -- (0, 1);
\draw (-1, 0) -- (0, 1);
   \draw[dashed, thick] (-2, -0.5) -- (-1, 0);
\draw[dashed, thick] (1, -0) -- (2, -0.5);
\draw[dashed, thick] (0, 1) -- (0, 2); 
   \node at (0, 1) {\textbullet};
   \node at (-1, 0) {\textbullet};
   \node at (1, 0) {\textbullet};
   \node at (0.3, 1.3) {$x_1$};
   \node at (1.5, -0) {$x_2$};
   \node at (-1.5, -0) {$x_3$};
   \node at (0, -0.25) {$y_{23}$};
   \node at (0.7, 0.7) {$y_{12}$};
   \node at (-0.7, 0.7) {$y_{31}$};
  \end{tikzpicture}
 \end{center}
 \label{fig:3pt-OneLoop-phi3}
 \caption{Triangle diagram in $\phi^3$ theory.}
\end{figure}
By using the algorithm described in section \ref{sec:review}, we can write the loop integrand for this diagram recursively
\begin{eqn}\label{A31integrand0}
&  \begin{tikzpicture}
   \draw (1, 0) -- (-1, 0);
\draw (1, 0) -- (0, 1);
\draw (-1, 0) -- (0, 1);
   \draw[dashed, thick] (-2, -0.5) -- (-1, 0);
\draw[dashed, thick] (1, -0) -- (2, -0.5);
\draw[dashed, thick] (0, 1) -- (0, 2); 
   \node at (0, 1) {\textbullet};
   \node at (-1, 0) {\textbullet};
   \node at (1, 0) {\textbullet};
    \node at (0.3, 1.3) {$x_1$};
   \node at (1.5, -0) {$x_2$};
   \node at (-1.5, -0) {$x_3$};
   \node at (0, -0.25) {$y_{23}$};
   \node at (0.7, 0.7) {$y_{12}$};
   \node at (-0.7, 0.7) {$y_{31}$};
  \end{tikzpicture}
  = \frac{1}{x_1 + x_2 + x_3} \Bigg[ L(x_1+ y_{12}, y_{13}, x_3, y_{23}, x_2 + y_{12}) \\
  &\hspace{3.2cm}+ L(x_3+ y_{23}, y_{13}, x_1, y_{12}, x_2 + y_{23})
  + L(x_1+ y_{13}, y_{12}, x_2, y_{23}, x_3 + y_{13})  \Bigg]~,
\end{eqn}
where the term $L(x_1, y_1, x_2, y_2, x_3)$ corresponds to the following diagram and is evaluated in \eqref{L-5point},
\begin{eqn*}
 L(x_1, y_1, x_2, y_2, x_3) 
 = \begin{tikzpicture}[baseline]
    \draw (-2, 0) -- (2, 0);
    \node at (-2, 0) {\textbullet};
    \node at (-2, -0.25) {$x_1$};
    \node at (-1, +0.25) {$y_1$};
    \node at (0, 0) {\textbullet};
    \node at (0, -0.25) {$x_2$};
    \node at (1, +0.25) {$y_2$};
    \node at (2, 0) {\textbullet};
    \node at (2, -0.25) {$x_3$};
   \end{tikzpicture}~.
\end{eqn*}
This enables us to write the integrand \eqref{A31integrand0} as a sum of three terms 
\begin{eqn}\label{A31integrand}
 A_3^{(1)}= A_{3, 1}^{(1)} + A_{3, 2}^{(1)} + A_{3, 3}^{(1)}~,
\end{eqn}
which are given as 
\begin{eqn}\label{3point-integrand}
 A_{3, 1}^{(1)}= \frac{1}{E_T} \frac{1}{E_T + 2 y_{23}}  \frac{1}{x_1 + y_{12} + y_{13}} \frac{1}{x_2 + y_{12} + y_{23}} \frac{1}{x_3 + y_{23} + y_{13}} &\Biggl\{ \frac{1}{x_1 + x_2 + y_{13} + y_{23} } \\
 & + \frac{1}{x_1 + x_3 + y_{12} + y_{23}} \Biggr\}~,\\
 A_{3, 2}^{(1)} = \frac{1}{E_T} \frac{1}{E_T+ 2 y_{13}} \frac{1}{x_1 + y_{12} + y_{13}} \frac{1}{x_2 + y_{12} + y_{23}} \frac{1}{x_3 + y_{23} + y_{13}} &\Biggl\{ \frac{1}{x_1 + x_2 + y_{13} + y_{23} } \\
 & + \frac{1}{x_2 + x_3 + y_{12} + y_{13}} \Biggr\}~,\\
 A_{3, 3}^{(1)}=\frac{1}{E_T} \frac{1}{E_T + 2 y_{12}} \frac{1}{x_1 + y_{12} + y_{13}} \frac{1}{x_2 + y_{12} + y_{23}} \frac{1}{x_3 + y_{23} + y_{13}} &\Biggl\{ \frac{1}{x_1 + x_3 + y_{12} + y_{23} }  \\
 & + \frac{1}{x_2 + x_3 + y_{12} + y_{13}} \Biggr\}~,
\end{eqn}
with $x_1 = |\bm k_1|$, $x_2 = |\bm k_2|$, $x_3 = |\bm k_3|$ and $y_{12} = y_{21} = |\bm l|$, $y_{23} = y_{32} = |\bm l + \bm k_2|$, $y_{31} = y_{13} = |\bm l - \bm k_1|$, and  $E_T = k_1 + k_2 + k_3$~. The value of the diagram is obtained after performing the integral over the loop momenta and is given as 
\be\label{3point-integral1}
\bm A_3^{(1)} = \int d^5 l\ A_3^{(1)}~.
\ee
It is difficult to perform this integral for a generic configuration of momenta. However, it is possible to evaluate them in the \textit{Squeezed Limit} (where we take $|\bm k_3| \to 0$). Such a limit was also useful in analyzing the 3-point function of CFTs in momentum space \cite{Bzowski:2013sza}. In this limit, the configuration of momenta becomes the following,
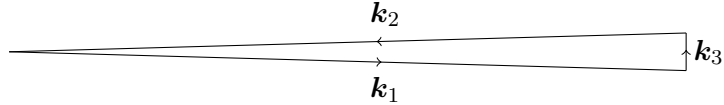
\begin{figure}[H]
 \centering 
 \begin{tikzpicture}
    \draw[fermionbar, black] (-7, 0) -- (2, 0.25);
    \draw[fermion, black] (-7, 0) -- (2, -0.25);
    \draw[fermion, black] (2, -0.25) -- (2, 0.25);
    \node at (-2, -0.5) {$\bm k_1$};
    \node at (-2, 0.5) {$\bm k_2$};
    \node at (2.3, 0) {$\bm k_3$};
 \end{tikzpicture}
\end{figure}
The integrands tremendously simplify in this limit and they all can be written in terms of the following three integrands, 
\begin{eqn}
 I_{3,1}^{(1)} &= \frac{1}{8E_T}\frac{1}{\big(k_2 + |\bm l + \bm k_2|\big)^2 } \frac{1}{\big( k_2 + l + |\bm k_2 + \bm l| \big)^2 } \frac{1}{|\bm k_2 + \bm l|}, \\
 I_{3,2}^{(1)} &= \frac{1}{4E_T} \frac{1}{k_2 + |\bm l + \bm k_2| } \frac{1}{\big( k_2 + l + |\bm k_2 + \bm l| \big)^3 } \frac{1}{|\bm k_2 + \bm l|}, \\
 I_{3, 3}^{(3)} &= \frac{1}{4E_T}  \frac{1}{k_2 + l} \frac{1}{\big( k_2 + l + |\bm k_2 + \bm l| \big)^3 } \frac{1}{|\bm k_2 + \bm l|}.
\end{eqn}
The original integrands in \eqref{3point-integrand} become
\begin{eqn}
A_{3,1}^{(1)} \overset{|\bm k_3| \to 0}{=\joinrel=} I_{3, 1}^{(1)} + I_{3, 2}^{(1)} , \qquad 
A_{3,2}^{(1)} \overset{|\bm k_3| \to 0}{=\joinrel=} I_{3, 1}^{(1)} + I_{3, 2}^{(1)} = A_{3, 1}^{(1)}, \qquad 
A_{3,3}^{(1)} \overset{|\bm k_3| \to 0}{=\joinrel=} 2I_{3, 3}^{(1)} ~,
\label{squeezed-itegrands}
\end{eqn}
and hence the integral \eqref{3point-integral1} can be expressed as
\be\label{3point-integral2}
\bm A_3^{(1)} = 2\int d^5 l \big[ A_{3,1}^{(1)}  + A_{3,2}^{(1)} + A_{3,3}^{(1)}\big]~. 
\ee
These integrals are similar to the integrals in the previous section and can be evaluated using the same methods. The final result is as follows
\begin{eqn}\label{triangle-squeezed}\boxed{
    \bm A_3^{(1)} = -\frac{S_3}{24 E_T} \big[ \pi^2 - 1 + 6 \log \frac{k_2}{\Lambda} \big]~.
}\end{eqn}
Since the three-point function is exactly known from the CFT side (see equation \eqref{CFT3pt}), it should also be possible to also evaluate it using Witten diagrams and check that the computations agree. As illustrated in section \ref{sec:Concl}, this implies that in a particular renormalization scheme, it is possible to find counterterms that cancel the divergences and lead to the structure of the expected CFT three-point function. Following the discussion in appendix \ref{app:3pt} it is expected that the momenta dependence of this diagram for generic momenta takes the form 
\be
\frac{1}{E_T}\sum_{\mbox{\footnotesize{perm}}[k_1, k_2, k_3]} \log \frac{E_T^2 }{k_1 \Lambda}
\ee
with $E_T = k_1 + k_2 + k_3$ and it would be interesting to derive this directly by evaluating the integrals in equation \eqref{3point-integral1} and to study this using the cosmological cutting rules \cite{Melville:2021lst} or the cosmological bootstrap \cite{Arkani-Hamed:2018kmz}. 

\subsection*{Box Diagrams}
In the previous section, we evaluated the triangle diagram in the squeezed limit. Using the same techniques one can also evaluate the box diagram in the squeezed/collinear limit when the adjacent or opposite legs are collinear, i.e., 
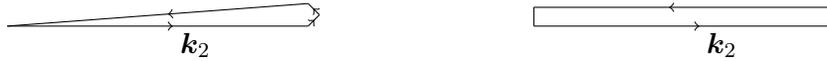
\begin{figure}[H]
\centering
\begin{tikzpicture}
\begin{scope}[shift={(-3.5, 0)}]
    \draw[fermion, black] (-4, 0) -- (0, 0);
    \draw[fermion, black] (0, 0) -- (0.15, 0.15);
    \draw[fermion, black] (0.15, 0.15) -- (0, 0.3);
    \draw[fermionbar, black] (-4, 0) -- (0, 0.3);
    \node at (-1.5, -0.25) {$\bm k_2$};
    \end{scope}
    \begin{scope}[shift={(3.5, 0)}]
    \draw[fermion, black] (-4, 0) -- (0, 0);
    \draw[fermion, black] (0, 0.25) -- (-4, 0.25);
    \draw (-4, 0.25) -- (-4, 0);
    \draw (0, 0.25) -- (0, 0);
    \node at (-1.5, -0.25) {$\bm k_2$};
    \end{scope}
\end{tikzpicture}
\caption{Configuration of momenta in the squeezed limit for the Box diagram.}
\end{figure}
After applying the algorithm described in the previous section, the loop integral can be simplified to the following form
\be
\frac{1}{E_T}\int \frac{d^5 l}{l^{a_1} (k_2 + l)^{a_2} ( k_2+ l + |\bm k_2 + \bm l|)^{a_3} }
\ee
with $a_1 + a_2 + a_3 = 6$ and $E_T$ is the total energy pole. These are similar to the integrals we have had until now. Although it is possible to do these integrals, it is not clear what it is useful for other than some consistency checks for CFT correlation functions in certain special limits (e.g., when two points become coincident).

\subsection{Extension to time-dependent interactions}\label{sec:wavefunc}
Until now we have considered theories where the interaction terms respected conformal invariance. For example, two main cases that we studied, had the following interactions: $\phi^4$ theory in $D= 4$ and $\phi^3$ theory in $D = 6$. However, the method that we discussed in the previous sections can also be used to work in other situations. For example,  $\phi^3$ theory in $D=4$. In this example, even the tree-level Witten diagrams have higher transcendentality than their conformal counterpart (which is $\phi^3$ theory in $D=6$). For example, we first review the computation of the  four-point function \cite{Arkani-Hamed:2015bza} at the tree level and then extend it to the one-loop case. 
\begin{figure}[H]
    \centering
    \begin{tikzpicture}
        \draw[thick, black] (-3, 0) -- (3,0); 
        \draw[fermion, black] (-2.5, 0) -- (-2, -2);
        \draw[fermion, black] (-1.5, 0) -- (-2, -2);
        \draw[fermion, black] (2.5, 0) -- (2, -2);
        \draw[fermion, black] (1.5, 0) -- (2, -2);
        \draw[fermion, black] (-2, -2) -- (2, -2);
        \node at (-2.6, -1) {$\bm k_1$};
        \node at (-1.4, -1) {$\bm k_2$};
        \node at (2.6, -1) {$\bm k_4$};
        \node at (1.4, -1) {$\bm k_3$};
    \end{tikzpicture}
    \caption{Tree level 4-point function in  $\phi^3$ theory for $D = 4$}
    \label{phi3tree-4d}
\end{figure}
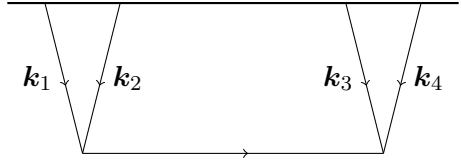
For this theory, in dS$_4$ space, the interaction vertices come with a factor of $\eta^{-1}$ (see the definition of $\lambda_k(\eta)$ in \eqref{action2}). Therefore the value of this diagram \ref{phi3tree-4d} in dS$_4$ is given as 
\be
\intnsinf \frac{d\eta_1}{\eta_1} \frac{d\eta_2}{\eta_2} e^{ik_{12}\eta_1}e^{ik_{34}\eta_2} G(\eta_1, \eta_2, |\bm k_1 + \bm k_2|)~.
\ee
where $k_{12} = |\bm k_1| + |\bm k_2|$ and $k_{34} = |\bm k_3| + |\bm k_4|$. The presence of factors $\eta_1^{-1} \eta_2^{-1}$ in the integration makes it different from the computations that have already been performed. However, it is possible to reformulate this calculation using the procedures outlined in section \ref{sec:recursion}. To do this we can recast $(\eta_1 \eta_2)^{-1}$ as an integral as shown below
\be
\int_{k_{12}}^\infty \int_{k_{34}}^{\infty}  \intnsinf d\eta_1 d\eta_2\, e^{i(k_1 
+ k_2)\eta_1}e^{i(k_3 + k_4)\eta_2} G(\eta_1, \eta_2, |\bm k_1 + \bm k_2|)~.
\ee
This integral is evaluated to be \cite{Arkani-Hamed:2015bza}, 
\be
\frac{1}{2k_I} \Big[ \text{Li}_2\frac{k_{12} -k_I}{k_{12} + k_{34}} + \text{Li}_2\frac{k_{34} - k_I}{k_{12} + k_{34}} + \log\frac{k_{12} + k_I}{k_{12} + k_{34}} \log\frac{k_{34} + k_I}{k_{12} + k_{34}} - \frac{\pi^2}{6} \Big]~,
\ee
with $k_I = |\bm k_1 + \bm k_2|$. The structural similarity between this expression and the one-loop diagram for $\phi^4$ theory in dS$_4$ \eqref{4ptoneloop} suggests that one might be able to derive these using other methods such as cosmological cutting rules \cite{Melville:2021lst}. It might be interesting to look at such connections and we postpone this to a future project.

\subsubsection*{One-Loop}
We now demonstrate how we can evaluate the contribution at one-loop for the figure \ref{fig:2pt-oneloop-phi3} for any general cosmology. The integrand there is of the form 
\be
\bm A^{4, (1)}_{\phi^3} = \intnsinf d\eta_1 d\eta_2 \frac{e^{i k \eta_1}}{\eta^\rho} \frac{e^{i k\eta_2}}{\eta^\rho} \int d^3l \, G(\eta_1, \eta_2, \bm l)G(\eta_1, \eta_2, |\bm l + \bm k|)~,
\ee
with $\rho$ being a positive number which is obtained from the vertex factors containing  $a(\eta) = \frac{1}{\eta^\rho}$. By using an integral transform we can now convert this integral to a form that we have evaluated before. This leads us to,
\be
\bm A^{4, (1)}_{\phi^3} = \intsinf \frac{ds_1 ds_2}{\Gamma[\rho]^2} (s_1 s_2)^{\rho -1} \int d^3l \times 
\begin{tikzpicture}[baseline]
    \draw (0, 0.1) circle (0.75);
    \node at (-0.75, 0.1) {\textbullet};
    \node at (0.75, 0.1) {\textbullet};
    \node at (-1.4,0.1) {\small{$k - i s_1$}};
    \node at (1.4,0.1) {\small{$k - i s_2$}};
    \node at (0, 1.1) {$y_1$};
    \node at (0, -0.9) {$y_2$};
\end{tikzpicture}~,
\ee
where $y_1 = |\bm l|$ and $y_2 = |\bm l + \bm k|$. The loop integral is exactly of the form \eqref{2pt-phi3-bubble-5dim} and leads to a similar expression. Although it is challenging to obtain a general form for the aforementioned integration, one can perform the integrals in analytic form for specific values of $\rho$ (with $\rho> 0$). For example, for $\rho = 1$ (which would be the case for $\phi^3$ theory in $dS_4$) the analytic structure of the integral only contains terms with transcendtality of $\log k$.

We also discuss another example where we have two kinds of interactions in the bulk, for example, $g \phi^3 + \lambda \phi^4 $ in dS$_4$. Since one interaction is conformal invariant we shall only have to extend the momenta for the second point as shown below. 
\begin{figure}[H]
\centering
 \begin{tikzpicture}
\draw (0,0) circle (1);
\draw[dashed, fermion, thick, black] (-1.5, 1) -- (-1, 0);
\draw[dashed, fermion, thick, black] (-1.5, -1) -- (-1, 0);
\draw[dashed, fermion, thick, black] (1, 0) -- (2, 0);
\node at (-1.6, 0.5) {$\bm k_1$};
\node at (-1.6, -0.5) {$\bm k_2$};
\node at (1.6, 0.25) {$\bm k_3$};
\node at (-1, -0) {\textbullet};
\node at (1, -0) {\textbullet};
\node at (-1.25,0) {${\scriptsize{x_1}}$};
\node at (0.75,0) {${\scriptsize{x_2}}$};
\node at (0, 1.25) {${\scriptsize{y_1}}$};
\node at (0, -1.25) {${\scriptsize{y_2}}$};
\end{tikzpicture}
\caption{One-Loop contribution to the Three-Point Function. }
\end{figure}
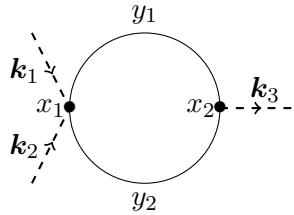
The contribution of this diagram is 
\be
\bm A_{\phi^4 + \phi^3} = \intsinf ds\int d^3l \times 
\begin{tikzpicture}[baseline]
    \draw (0, 0.1) circle (0.75);
    \node at (-0.75, 0.1) {\textbullet};
    \node at (0.75, 0.1) {\textbullet};
    \node at (-1.1,0.1) {\small{$k_{12} $}};
    \node at (1.4,0.1) {\small{$k_3 - i s$}};
    \node at (0, 1.1) {$y_1$};
    \node at (0, -0.9) {$y_2$};
\end{tikzpicture}~.
\ee
This integration can be explicitly performed and gives an answer very similar to equation \eqref{4ptoneloop} but with different coefficients\footnote{For the convenience of the reader, we have attached a Mathematica notebook that computes this integral. }.

\section{Conclusion and Discussion}\label{sec:Concl}
\subsection*{\normalfont\textit{Summary}}\label{sec:summary}
In this paper, we have studied the wave function coefficients for conformally coupled scalars at one and two-loop level. These results can also be used to compute the cosmological correlation functions for any general cosmology and transition amplitudes in AdS for conformally coupled scalars under suitable assumptions. The use of recursion relations, developed in \cite{Arkani-Hamed:2017fdk}, has been a key technique in simplifying the loop integrands, enabling us to evaluate these integrals in a more convenient form. Some of the diagrams which are exactly calculable using these methods include Cactus, Sunset, and Bubble, at two point, with a possible extension to similar diagrams at three and four-point. It is also possible to evaluate the triangle diagram in the squeezed limit (see section \ref{sec:triangle}) using these methods and also bootstrap its general structure. In particular, the cactus diagram, which naively gives a divergent expression from the bulk integrals can be regularized in a consistent manner. While our results have primarily focused on self-interacting conformally coupled scalar fields, they can be easily extended to any cosmological model, including cases where the interaction potential is not conformally invariant, as discussed in section \ref{sec:wavefunc}.

Inspired by \cite{Albayrak:2020isk}, we adopted the hard cutoff regularization scheme for the loop integrals. While this approach may be suitable for scalar fields, it is not an appropriate technique for gauge theories as this procedure is not gauge invariant.  Compared to other regularization methods, dimensional regularization presents more challenges as the form of the propagators in momentum space (which typically are Bessel functions) is dependent on the dimensions of spacetime. As a result, it becomes difficult to analytically continue the method to arbitrary dimensions.

In this paper, we have evaluated loop integrals for scalar fields having a particular scaling dimension with time-dependent interactions. However, it is important to note that our approach has been limited to the regularization of the expressions. We have not yet performed the renormalization of the diagrams. These need to be carefully handled and have been discussed in several works in various contexts \cite{Bertan:2018khc, Banados:2022nhj, Lee:2023jby} and one of the next steps would be to apply these to our construction. There has also been some discussion on the choice of the regulator in \cite{Katsianis:2019hhg, Katsianis:2020hzd} and it would be interesting to also perform the loop integrals in our paper using these regularization procedures.

\subsection*{\normalfont\textit{Renormalization}}\label{sec:renorm}
Although we have not discussed this explicitly in the bulk of the paper, it is possible to  follow the usual schemes for renormalizing the diagrams. It should be possible to add the necessary counterterms in a way such that they preserve the conformal symmetry of these correlators. 
This would allow one to read off the anomalous dimension of the theory from the coefficient of $\log k$ terms in the case of 2-point functions (see equation \eqref{anomalous}) that appear when the CFT correlation functions in momentum space are expanded about their conformal dimension, $\Delta = \frac{d-1}{2}$. A similar argument also works for the 3-point function. This renormalization scheme has been discussed at one-loop in momentum space for two point functions in $\phi^4$ theory in dS$_4$ \cite{Gorbenko:2019rza} and there has been some progress in utilizing such an approach in position space as well, for example, \cite{Bertan:2018khc, Banados:2022nhj}. It would be interesting to explicitly compute the counterterms and use them to study the renormalization of higher-point functions in momentum space.

\subsection*{\normalfont\textit{Relation with Transition Amplitudes}}\label{app:transamp}
An object that is closely related to the wave function coefficients are known as \textit{transition amplitudes} in AdS and is defined as \cite{Balasubramanian:1998de, Balasubramanian:1999ri, Raju:2011mp} 
\be
T(\bm k_{ij}) (2\pi)^d \delta^d\big( \sum_{ij} \bm k_{ij}  \big) = \braket{s|O(\bm k_{3j_1} \cdots O(\bm k_{3j_n}) )|s'}
\ee
with the momentum-conserving delta function being extracted out on the left-hand side. The states $\ket{s}$, $\ket{s'}$\footnote{These are coherent states as they solve the classical equations of motion in the bulk. We remind the reader that the normalizable modes in the bulk are dual to states in the CFT and the non-normalizable modes are dual to sources.} are labeled by the momenta $\bm k_{1j}$ and $\bm k_{2j}$ respectively and are dual to linear combinations of normalizable modes. These are evaluated in the same way as one evaluates the wave function coefficients, i.e.,  the bulk-boundary correlators are chosen to be the normalizable modes and the rest of the diagram is exactly similar, where one draws the usual bulk-bulk propagators and contracts them with the normalizable bulk-boundary propagators. 
To compare with the cases considered in this paper, all the non-normalizable modes are switched off. Hence all the results in this paper can also be extended to AdS by a suitable analytic continuation \cite{Maldacena:2002vr}.

\subsection*{\normalfont\textit{Minimally Coupled Scalars}}
By comparing the propagators in minimally and conformally coupled scalars, we see that they have very similar structures. The former comprises of  the following Bessel functions $J_{3/2}, K_{3/2}$ (in $D = 4$), whereas the latter has the Bessel functions $J_{1/2}, K_{1/2}$. Since the Bessel functions $J_{3/2}, K_{3/2}$ almost follow the Cauchy's exponential functional equation, it might be possible to extend the algorithm for the recursion relations reviewed in section \ref{sec:recursion} to minimally coupled scalars as well.

\subsection*{\normalfont\textit{Spinning Loops}}
The recursion relations developed in \cite{Arkani-Hamed:2017fdk} can be extended to  Yang-Mills theory at tree level \cite{Albayrak:2018tam, Albayrak:2019asr}. This allows us to expand the integrand for loop diagrams \cite{Albayrak:2020bso} in partial fractions.  Therefore many diagrams discussed in this paper can also be evaluated in Yang-Mills theory after regularizing them in a gauge invariant fashion. It would be interesting to evaluate these diagrams in the self-dual sector of Yang-Mills as well. Since there is evidence of double copy working at tree level for this theory \cite{Lipstein:2023pih}, one can also try to evaluate them at loop level for gravity and check if the double copy relations still hold.

\subsubsection*{Acknowledgements}
We sincerely thank Dean Carmi, Victor Godet,  Dileep Jatkar, Savan Kharel, Alok Laddha, Arthur Lipstein, Raghu Mahajan, Suvrat Raju, Ashoke Sen, Kostas Skenderis, Tom Westerdijk, and all members of the ICTS and HRI string group for several useful discussions and very helpful feedback. Research at ICTS-TIFR is supported by the Department of Atomic Energy, Government of India, under Project Identification No. RTI4001. KS is supported by Infosys Fellowship at HRI, India. KS would like to thank ICTS, Bengaluru for hospitality where a part of this work was completed. CC would like to thank ICTP, Trieste for hospitality where a part of this work was completed, and would like to thank the participants of the workshop on Scattering Amplitudes and Cosmology for several useful discussions.  Both authors would like to acknowledge the hosts of Kavli Asian Winter School 2023 held in IBS Daejong, where this project was initiated. 

\begin{appendix}

\section{Regularization of 2-loop Cactus Diagram}\label{app:cactus2}
\begin{figure}[H]
\centering
    \begin{tikzpicture}
\begin{scope}[shift={(2, 0)}]
\draw[fermion, black] (0,0.5) circle (0.5);
    \draw[fermion, black] (0,1.5) circle (0.5);
\node at (0, 0) {\textbullet};
\node at (0, -0.25) {$x_1$};
\draw[dashed, thick] (-1, -0) -- (0, -0);
\draw[thick, fermion, dashed, black] (0, -0) -- (1, -0);
\node at (0.5, -0.25) {$\bm k$};
\node at (0, 1) {\textbullet};
\node at (0, 0.75) {$x_2$};
\node at (-0.8, 0.5) {$y_1$};
\node at (0.8, 0.5) {$y_1$};
\node at (-0.8, 1.5) {$y_2$};
\end{scope}
\end{tikzpicture}
\caption{2-loop Cactus diagram }
\label{fig:2cactus}
\end{figure}
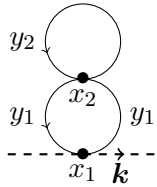
\be
\ee
The expression for the 2-loop cactus diagram is given as 
\be
R_{(2)} = \intnsinf d\eta_1 d\eta_2\, e^{i x_1 \eta_1} \big[G(\eta_1, \eta_2, y_1)\big]^2 G(\eta_2, \eta_2, y_2)
\ee
where $x_1 = 2 |\bm k|$, $y_1 = |\bm l_1|$ and $y_2 = |\bm l_2|$. This is naively divergent because of the lack of the damping factor in the $\eta_2$ integral. Since the flat space limit of the integral is finite, we expect that this divergence can be regularized. To do this, we re-write the integral as suggested in figure \ref{fig:2cactus} by inserting a regulator $e^{i x_2 \eta_2}$ in the integrand (we suppress the limits of the integrals from this step),
\be
R_{(2)}  = \lim_{x_2 \to 0}\hat R_{(2)} \equiv\lim_{x_2 \to 0} \int d\eta_1 d\eta_2 e^{i x_1 \eta_1} e^{i x_2 \eta_2} \big[G(\eta_1, \eta_2, y_1)\big]^2 G(\eta_2, \eta_2, y_2)
\ee
Note that adding this factor is not enough by itself to give a finite answer for the integral as  will be demonstrated in the steps below. However, this does allow for some integrals in the intermediate steps to be finite. For example, the integral $\intnsinf d\eta $ is divergent, however, by inserting the regulator $e^{i x \eta}$ in the integrand, we get a finite answer in the intermediate step $\intnsinf d\eta e^{i x \eta} = \frac{1}{i x}$~. With this regulator, we have an integral of the form in equation \eqref{nimabasis} and can therefore use the algorithm of section \ref{sec:recursion}
\be
\hat R_{(2)} = \int d\eta_1 d\eta_2 e^{i x_1 \eta_1} e^{i x_2 \eta_2} \big[G(\eta_1, \eta_2, y_1)\big]^2 G(\eta_2, \eta_2, y_2)~. 
\ee
Due to the presence of the bulk-bulk Green's function at a coincident point, this expression is a bit tricky to write using the recursion relation directly and hence we evaluate it step by step. First, we insert the time translation operator in the integrand and use the fact that the integrand is zero at the boundary to give us the following relation
\be
\int d\eta_1 d\eta_2 (-i) \left(\frac{\p}{\p \eta_1} + \frac{\p}{\p \eta_2} \right) \Bigg[ e^{i x_1 \eta_1} e^{i x_2 \eta_2} \big[G(\eta_1, \eta_2, y_1)\big]^2 G(\eta_2, \eta_2, y_2) \Bigg] = 0~. 
\ee
The action of the derivatives on the three terms is given as 
\begin{enumerate}
    \item $e^{i x_1 \eta_1} e^{i x_2 \eta_2}$ : 
\be
\int d\eta_1 d\eta_2 (-i)  \left(\frac{\p}{\p \eta_1} + \frac{\p}{\p \eta_2} \right) \Bigg[ e^{i x_1 \eta_1} e^{i x_2 \eta_2}\Bigg]  \big[G(\eta_1, \eta_2, y_1)\big]^2 G(\eta_2, \eta_2, y_2) = \big( \eta_1 +\eta_2 \big) \hat R_{(2)}
\ee
\item $\big[G(\eta_1, \eta_2, y_1)\big]^2$ :
\begin{eqn}
    &\int d\eta_1 d\eta_2  e^{i x_1 \eta_1} e^{i x_2 \eta_2} (-i) \left(\frac{\p}{\p \eta_1} + \frac{\p}{\p \eta_2} \right)  \big[G(\eta_1, \eta_2, y_1)\big]^2 G(\eta_2, \eta_2, y_2) \\
    &= 
       - 2 \int d\eta_1 d\eta_2  e^{i x_1 \eta_1} e^{i x_2 \eta_2}  G(\eta_1, \eta_2, y_1) e^{i y_1 (\eta_1 + \eta_2)}  G(\eta_2, \eta_2, y_2) ~,
\end{eqn}
where this follows from \eqref{Deltasimp}. This integral is diagrammatically expressed as
\begin{figure}[H]
\centering
\begin{tikzpicture}
\node at (0,0) {\textbullet};
\node at (0, 1) {\textbullet};
    \draw[black] (0,1.5) 
    circle (0.5);
    \draw (0,0) -- (0,1);
    \node at (0, -0.25) {$x_1 + y_1$};
    \node at (0.75, 0.75) {$x_2 + y_1$};
    \node at (-0.8, 1.5) {$y_2$};
    \node at (-0.25, 0.5) {$y_1$};
\end{tikzpicture}
\end{figure}

\item $ G(\eta_2, \eta_2, y_2)$ :
\begin{eqn}\label{2Cactus-term3}
&\int d\eta_1 d\eta_2  e^{i x_1 \eta_1} e^{i x_2 \eta_2}  \big[G(\eta_1, \eta_2, y_1)\big]^2  (-i)\left(\frac{\p}{\p \eta_1} + \frac{\p}{\p \eta_2} \right)  G(\eta_2, \eta_2, y_2) \\
&= 
\int d\eta_1 d\eta_2  e^{i x_1 \eta_1} e^{i x_2 \eta_2}  \big[G(\eta_1, \eta_2, y_1)\big]^2 (-i) \left( \frac{\p}{\p \eta_2} \right)  G(\eta_2, \eta_2, y_2) ~.
\end{eqn}
This is the term that naively leads to a divergent integral because of the coinciding Theta functions. However, we can regulate it using point-splitting regularization\footnote{The correct normalization is decided by spectral representation for the bulk-bulk Green's function given in equation \eqref{spectral}.}, i.e., 
\be\label{regularapp}
(-i)\frac{\p}{\p \eta_2} G(\eta_2, \eta_2, y_2) = (-i) \lim_{\eta'_2 \to \eta_2} \left( \frac{\p}{\p \eta_2} + \frac{\p}{\p \eta'_2} \right) G(\eta_2, \eta'_2, y_2) = - e^{2 i y_2 \eta_2}
\ee
The main difference between the 1-Cactus and the 2-Cactus is that it was possible to derive the former without using this method as shown in equation \eqref{spectral}. However, following the same method, in this case, leads to a divergent expression. Thus, the algorithm provides a natural way to regulate such divergences. Using this regularized expression, equation \eqref{2Cactus-term3} becomes,
\begin{eqn}
&\int d\eta_1 d\eta_2  e^{i x_1 \eta_1} e^{i x_2 \eta_2}  \big[G(\eta_1, \eta_2, y_1)\big]^2 \left(\frac{\p}{\p \eta_1} + \frac{\p}{\p \eta_2} \right)  G(\eta_2, \eta_2, y_2) \\
&=- \int d\eta_1 d\eta_2  e^{i x_1 \eta_1} e^{i (x_2 + 2 y_2) \eta_2}  \big[G(\eta_1, \eta_2, y_1)\big]^2 
\end{eqn}
The diagrammatic representation for this integral is given below
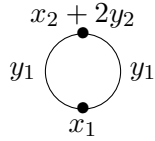
\begin{figure}[H]
\centering
\begin{tikzpicture}
\node at (0,0) {\textbullet};
\node at (0, 1) {\textbullet};
    \draw (0,0.5) 
    circle (0.5);
    \node at (0, -0.25) {$x_1 $};
    \node at (0, 1.25) {$x_2 + 2y_2$}; 
    \node at (-0.8, 0.5) {$y_1$};
    \node at (0.8, 0.5) {$y_1$};
\end{tikzpicture}
\end{figure}
This diagram is not to be confused with the usual bubble diagram as the two internal propagators in this one carry the same momenta.
\end{enumerate}

Combining the expressions above, we end up with the following diagrammatic expression for $\hat R_{(2)}$
\begin{eqn}
    \hat R_{(2)} =&\frac{1}{x_1 + x_2}
    \Big[2 \begin{tikzpicture}[baseline]
    \begin{scope}[shift={(0, 0)}]
\node at (0,0) {\textbullet};
\node at (0, 1) {\textbullet};
    \draw[black] (0,1.5) 
    circle (0.5);
    \draw (0,0) -- (0,1);
    \node at (0, -0.25) {$x_1 + y_1$};
    \node at (0.75, 0.75) {$x_2 + y_1$};
    \node at (-0.8, 1.5) {$y_2$};
    \node at (-0.25, 0.5) {$y_1$};
\end{scope}
\end{tikzpicture}
+ 
\begin{tikzpicture}[baseline]
\begin{scope}[shift={(4, 0)}]
    \node at (0,0) {\textbullet};
\node at (0, 1) {\textbullet};
    \draw[black] (0,0.5) 
    circle (0.5);
    \node at (0, -0.25) {$x_1 $};
    \node at (0, 1.25) {$x_2 + 2y_2$}; 
    \node at (-0.8, 0.5) {$y_1$};
    \node at (0.8, 0.5) {$y_1$};
\end{scope}    
  \end{tikzpicture}
  \Big]
\end{eqn}
  The two diagrams can be evaluated in a similar manner as above using the regularization described in equation \eqref{regularapp}.
\begin{enumerate}
    \item Diagram 1 :
    \begin{eqn}
    \begin{tikzpicture}[baseline]
    \node at (0,0) {\textbullet};
\node at (0, 1) {\textbullet};
    \draw[black] (0,1.5) 
    circle (0.5);
    \draw (0,0) -- (0,1);
    \node at (0, -0.25) {$x_1 + y_1$};
    \node at (0.75, 0.75) {$x_2 + y_1$};
    \node at (-0.8, 1.5) {$y_2$};
    \node at (-0.25, 0.5) {$y_1$};
    \end{tikzpicture} 
    &= \frac{1}{(x_2 + x_1 + 2y_1 )} \Big[
     \begin{tikzpicture}[baseline]
    \node at (0,0) {\textbullet};
\node at (0, 1) {\textbullet};
    \draw[black] (0,1.5) 
    circle (0.5);
    
    \node at (0, -0.25) {$x_1 + 2y_1$};
    \node at (0, 0.75) {$x_2 + 2y_1$};
    \node at (-0.8, 1.5) {$y_2$};
    \end{tikzpicture}
    +
     \begin{tikzpicture}[baseline]
    \node at (0,0) {\textbullet};
\node at (0, 1) {\textbullet};
    
    \draw (0,0) -- (0,1);
    \node at (0, -0.25) {$x_1 + y_1$};
    \node at (0, 1.25) {$x_2 + y_1 + 2 y_2$};
    \node at (0.25, 0.5) {$y_1$};
    \end{tikzpicture}
    \Big]\\
    &= \frac{1}{(x_1 + x_2 + 2 y_1)} \Big[ \frac{1}{(x_2 + 2 y_1)(x_2 + 2y_2 + 2 y_1)} \frac{1}{x_1 + 2 y_1}\\
    &\qquad\qquad\qquad\qquad+ \frac{1}{ (x_1 + x_2 + 2 y_1 + 2 y_2)(x_1 + 2y_1)(x_2 + 2y_1 + 2 y_2) }  \Big]
    \end{eqn}

    \item Diagram 2 :
    \begin{eqn}
    \begin{tikzpicture}[baseline]
 \node at (0,0) {\textbullet};
\node at (0, 1) {\textbullet};
    \draw[black] (0,0.5) 
    circle (0.5);
    \node at (0, -0.25) {$x_1 $};
    \node at (0, 1.25) {$x_2 + 2y_2$}; 
    \node at (-0.8, 0.5) {$y_1$};
    \node at (0.8, 0.5) {$y_1$};
    \end{tikzpicture} 
    &= \frac{2}{x_1 + x_2 + 2 y_2} 
    \begin{tikzpicture}[baseline]
        \draw (0,0) -- (0,1);
        \node at (0,0) {\textbullet};
        \node at (0,1) {\textbullet};
        \node at (0.25, 0.5) {$y_1$};
        \node at (0, -0.25) {$x_1 + y_1$};
        \node at (0, 1.25) {$x_2 + y_1 + 2 y_2$};
    \end{tikzpicture}\\
    &= \frac{2}{(x_1 + x_2 + 2 y_2)} \frac{1}{(x_1 + x_2 + 2 y_1 + 2 y_2)(x_1 + 2y_1)(x_2 + 2y_1 + 2 y_2) } 
    \end{eqn}
\end{enumerate}

Combining everything we have an expression for $\hat R_{(2)}$ 
\begin{eqn}
    \hat R_{(2)} &= \frac{2}{x_1 + x_2} \Bigg\{ \frac{1}{(x_1 + x_2 + 2 y_1)} \Big[ \frac{1}{(x_2 + 2 y_1)(x_2 + 2y_2 + 2 y_1)(x_1 + 2 y_1)}\\
    &\hspace{5cm}+ \frac{1}{ (x_1 + x_2 + 2 y_1 + 2 y_2)(x_1 + 2y_1)(x_2 + 2y_1 + 2 y_2) }  \Big] \\
    &\hspace{2.75cm} + \frac{1}{(x_1 + x_2 + 2 y_2)(x_1 + x_2 + 2 y_1 + 2 y_2)(x_1 + 2y_1)(x_2 + 2y_1 + 2 y_2) }\Bigg\}
\end{eqn}

Clearly this expression is finite as we take $x_2 \to 0$ and therefore obtain the regulated expression for $R_{(2)} = \lim_{x_2 \to 0} \hat R_{(2)}$
\begin{eqn}
R_{(2)} &=  \frac{2}{x_1(x_1 + 2 y_1)} \Big[ \frac{1}{( 2 y_1)(2y_2 + 2 y_1)(x_1 + 2 y_1)}
+ \frac{1}{ (x_1 + 2 y_1 + 2 y_2)(x_1 + 2y_1)( 2y_1 + 2 y_2) }  \Big] \\
    &\qquad + \frac{1}{x_1(x_1  + 2 y_2)(x_1 + 2 y_1 + 2 y_2)(x_1 + 2y_1)( 2y_1 + 2 y_2) } ~.
\end{eqn}

Hence the two-loop cactus diagram is given as 
\be\label{cactusloopintegrand}
\bm R_{(2)} =\int  \frac{2d^3 l_1 d^3 l_2}{x_1(x_1 + 2 y_1)} \Big[ \frac{1}{( 2 y_1)(2y_2 + 2 y_1)(x_1 + 2 y_1)}
+ \frac{2}{ (x_1 + 2 y_1 + 2 y_2)(x_1 + 2y_1)( 2y_1 + 2 y_2) }  \Big] ~.
\ee
with $x_1 = 2|\bm k|$, $y_1 = |\bm l_1|$ and $y_2 = |\bm l_2|$~.

\subsection*{$n$-Cactus and other Tadpoles}
The regularization procedure described above can be used to evaluate the n-Cactus diagram as well.
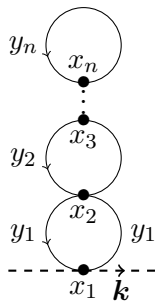
\begin{figure}[H]
\centering
    \begin{tikzpicture}
\begin{scope}[shift={(2, 0)}]
\draw[fermion, black] (0,0.5) circle (0.5);
\draw[fermion, black] (0,1.5) circle (0.5);
\draw[fermion, black] (0,3) circle (0.5);
\node at (0, 0) {\textbullet};
\node at (0, -0.25) {$x_1$};
\draw[dashed, thick] (-1, -0) -- (0, -0);
\draw[thick, dashed, fermion, black] (0, -0) -- (1, -0);
\node at (0.5, -0.25) {$\bm k$};
\node at (0, 1) {\textbullet};
\node at (0, 0.75) {$x_2$};
\node at (-0.8, 0.5) {$y_1$};
\node at (0.8, 0.5) {$y_1$};
\node at (-0.8, 1.5) {$y_2$};
\node at (0,2.35)  {$\vdots$};
\node at (-0.8, 3) {$y_n$};
\node at (0, 2) {$\bullet$};
\node at (0, 1.75) {$x_3$};
\node at (0, 2.5) {$\bullet$};
\node at (0, 2.75) {$x_n$};
\end{scope}
\end{tikzpicture}
\caption{n-loop Cactus diagram}
\end{figure}
The integral over $\eta_n$ is divergent but it is easy to see that it can be regulated in the same way as the 2-cactus diagram. First, one introduces a term $e^{i x_n \eta_n}$ in the integrand. Then, using the regularization technique used in equation \eqref{regularapp} we can regularize the $\eta_n$ integral. This process can then be repeated for the integrals after that. Using this, we can diagnose the divergences encountered in the loop integrand for any diagram of the kind 
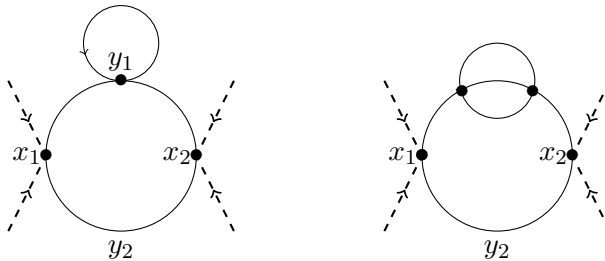
\begin{figure}[H]
\centering
 \begin{tikzpicture}
 \begin{scope}[shift={(-2, 0)}]
\draw (0,0) circle (1);
\draw[dashed, fermion, thick, black] (-1.5, 1) -- (-1, 0);
\draw[dashed, fermion, thick, black] (-1.5, -1) -- (-1, 0);
\draw[dashed, fermion, thick, black] (1.5, 1) -- (1, 0);
\draw[dashed, fermion, thick, black] (1.5, -1) -- (1, 0);

\node at (-1, -0) {\textbullet};
\node at (1, -0) {\textbullet};
\node at (0, 1) {\textbullet};
\draw[fermion, black] (0,1.5) circle (0.5);

\node at (-1.25,0) {${\scriptsize{x_1}}$};
\node at (0.75,0) {${\scriptsize{x_2}}$};
\node at (0, 1.25) {${\scriptsize{y_1}}$};
\node at (0, -1.25) {${\scriptsize{y_2}}$};
\end{scope}

\begin{scope}[shift={(3, 0)}]
\draw (0,0) circle (1);
\draw[dashed, fermion, thick, black] (-1.5, 1) -- (-1, 0);
\draw[dashed, fermion, thick, black] (-1.5, -1) -- (-1, 0);
\draw[dashed, fermion, thick, black] (1.5, 1) -- (1, 0);
\draw[dashed, fermion, thick, black] (1.5, -1) -- (1, 0);

\node at (-1, -0) {\textbullet};
\node at (1, -0) {\textbullet};
\node at (-1, -0) {\textbullet};
\node at (0.47, 0.85) {\textbullet};
\node at (-0.47, 0.85) {\textbullet};
\draw[fermion, black] (0,1.0) circle (0.5);

\node at (-1.25,0) {${\scriptsize{x_1}}$};
\node at (0.75,0) {${\scriptsize{x_2}}$};
\node at (0, -1.25) {${\scriptsize{y_2}}$};
\end{scope}
\end{tikzpicture}
\end{figure}
This resolves the issue of the tadpoles, cactus, and other diagrams with propagators purely lying in the bulk being divergent.

\section{Evaluating the Triangle Diagram for $k_3 \neq 0$}\label{app:k3neq0}
In section \ref{sec:triangle} we evaluated the triangle diagram in the squeezed limit. Here we present a perturbative approach to obtain a more general result away from the squeezed limit. 
\begin{figure}
  \centering
  \begin{tikzpicture}[x={(-0.6cm,-0.4cm)}, y={(1cm,0cm)}, z={(0cm,1cm)}]
    \draw[->] (0,0,0) -- (5,0,0) node[below]{$x$};
    \draw[->] (0,0,0) -- (0,5,0) node[right]{$y$};
    \draw[->] (0,0,0) -- (0,0,5) node[above]{$z$};
    
    \draw[->, thick] (0,0,0) -- (0,0,4);
    \node at (0,0.35,3) {$\bm {k_2}$};
    \draw[->, thick] (0,0,4) -- (0,-1,3.5);
    \node at (0,-0.5,4.25) {$\bm {k_3}$};
    \draw[->, thick] (0,0,0) -- (0,-1,3.5);
    \node at (0,-1.5,2.75) {$-\bm {k_1}$};
    \draw[->, thick] (0,0,0) -- (5,5,5);
    \node at (3.5,3.45,2.75) {$\bm {l}$};
    
    \node at (0,-0.3,1.8) {\small{$\psi$}};
    \draw (0,0,1.5) -- (0, -0.4, 1.45);
\draw (0,0,0.5) -- (0, 0.25, 0.4);
    \node at (0.8,0.7,1) {$\theta$};
\end{tikzpicture}
  \caption{Vector configuration of momenta in triangle diagram. We use the fact that the three external momenta $\bm k_1, \bm k_2, \bm k_3$ satisfy momentum conservation and therefore form a plane. We align this with the $y-z$ plane of the loop integral $\bm l$ with the vector $\bm k_2$ along the $z$-axis.}
  \label{fig:3d_coordinates}
\end{figure}
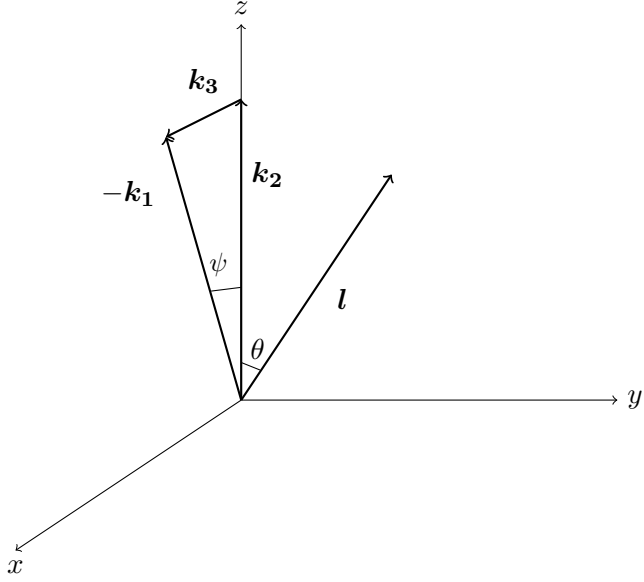
In the squeezed limit, we considered $k_1=k_2=k$ and $k_3=0$. However, a more general result for the one-loop triangle diagram can be derived by performing a series expansion in two parameters: $k_1=k_2+\epsilon_1$ and $k_3=\epsilon_2$. In this expansion, the loop propagators are affected since $|\bm l-\bm k_1| \neq |\bm l+\bm k_2|$. On the other hand, $|\bm l+\bm k_2|$ remains the same as before. To determine $|\bm l-\bm k_1|$, we utilize momentum configuration shown in Figure \ref{fig:3d_coordinates}, which is  $\bm k_1+\bm k_2+\bm k_3=0$.

For simplicity, we will work at $O(\epsilon_1^0)$\footnote{We provide a Mathematica notebook to generalize to higher orders.}, which implies $k_1=k=k_2$.
\begin{equation}
    |\bm l-\bm k_1|=|\bm l+\bm k_2+\bm k_3|=\sqrt{l^2+k^2+k_3^2+2 l k\cos{\q}+2 \bm k_2.\bm k_3+2\bm l.\bm k_3}\,,
\end{equation}
\begin{equation}
     \bm k_2= k\, \hat{\bm l}_z\,,\qquad \bm k_3= k_3 \cos(\pi+\frac{\psi}{2})\, \hat{\bm l}_y + k_3 \cos(\frac{\pi}{2}+\frac{\psi}{2})\, \hat{\bm l}_z\,,
\end{equation}
\begin{equation}
    \bm l= l \sin\theta \cos\phi\, \hat{\bm l}_x+l \sin\theta \sin\phi\, \hat{\bm l}_y + l \cos\theta\, \hat{\bm l}_z\,
\end{equation}
here $\psi$ is the angle between $\bm k_1$ and $\bm k_2$, it is easy to see that,
\begin{equation}
   \bm k_2.\bm k_3=-k k_3 \sin \left(\frac{\psi }{2}\right)\,,
 \end{equation}
 \begin{equation}
     \bm l.\bm k_3= -l k_3  \left(\cos (\theta ) \sin \left(\frac{\psi }{2}\right)+\sin (\theta ) \cos \left(\frac{\psi }{2}\right) \sin (\phi )\right)\,.
 \end{equation}
Using $\bm k_1+\bm k_2+\bm k_3=0$, we get $\sin(\psi/2)=k_3/2k$ and hence  
\begin{align}
    |\bm l-\bm k_1| =\sqrt{l^2+k^2+2 k l \cos\theta -2 k_3 l \left(\sqrt{1-\frac{k_3^2}{4 k^2}} \sin\theta \sin\phi+\frac{k_3}{2 k} \cos\theta \right)}  \,.
    \label{l-k1}
\end{align}
For example, we will work out the following integrand
\begin{align}
\begin{split}
   A_{3, 2}^{(1)}& = \frac{1}{E_T} \frac{1}{E_T+ 2 |\bm l-\bm{k_1}|} \frac{1}{k_1 + l + |\bm{l}-\bm{k_1}|} \frac{1}{k_2 + l + |\bm{l}+\bm{k_2}|} \frac{1}{k_3 + |\bm{l}+\bm{k_2}| + |\bm{l}-\bm{k_1}|}\\
 & \times \Biggl\{ \frac{1}{k_1 + k_2 +|\bm{l}-\bm{k_1}| +|\bm{l}+\bm{k_2}| } + \frac{1}{k_2 + k_3 + l + |\bm{l}+\bm{k_2}|} \Biggr\}~,\\
\end{split}
\label{gen}
\end{align}

We consider a series expansion around $k_3=0$. We substitute equation \eqref{l-k1} into equation \eqref{gen} and leave the $E_T$ factor out from this substitution to ensure the correct flat space-limit. This is also a natural thing to do from the point of view of the recursion relation described in section \ref{sec:recursion} as the overall factor of $E_T$ always appears in the LHS of the recursion relation.  The series can be written as follows:
\begin{equation}
     A_{3, 2}^{(1)}= \sum_{n=0}^{\infty} k_3^{n}\left(A_{3, 2}^{(1)}\right)_{n}\,,
     \label{pertse}
\end{equation}
here $\left(A_{3, 2}^{(1)}\right)_{0}$ is the integrand in squeezed limit as given in \eqref{squeezed-itegrands}. 

As we progress in the order of expansion, we observe a reduction in the level of divergence. This can be attributed to the structure of the propagators in the denominator of equation \eqref{gen}. Each additional power of $k_3$ in the numerator introduces an extra power of the loop momenta $l$ in the denominator. Consequently, beyond a certain order, the terms in the series become free from divergences.

Specifically, in the $\phi^{4}$ theory with $D=d+1=4$\footnote{To mimic the triangle diagram in the $\phi^3$ theory in $D = 6$, which is a three-point function, we will evaluate the triangle diagram for $\phi^4$ theory in $D = 4$ with collinear momenta at each vertex, which effectively reduces a six-point function of single trace operators to a three-point function of double trace operators. However, it is trivial to generalize this using the Mathematica code supplied with the submission. }, we have shown that the zeroth-order amplitude in $k_3$ and all higher-order amplitudes are finite. Similarly, in the $\phi^{3}$ theory with $D=d+1=6$, the first-order and all subsequent order amplitude are also finite (the zeroth order result is given in \eqref{triangle-squeezed}).

For example, the $O(k_3^1)$ integrand in the series \eqref{pertse} is.
\begin{align}
   \begin{split}
\frac{E_T}{k_3}  \left(A_{3, 2}^{(1)}\right)_{1} &=-\frac{2\pi k l^3 \left((5 \cos \theta +3) \sqrt{k^2+2 k l \cos \theta +l^2}+l (4 \cos \theta+6)\right)}{4 \left(k^2+2 k l \cos \theta+l^2\right) \left(\sqrt{k^2+2 k l \cos \theta+l^2}+k\right)^3 \left(\sqrt{k^2+2 k l \cos \theta +l^2}+k+l\right)^4}\\
  &-\frac{\pi  l^2 \left(13 k^3+k^2 \left(13 \sqrt{k^2+2 k l \cos \theta+l^2}+l (23 \cos \theta+6)\right)\right)}{4 \left(k^2+2 k l \cos \theta +l^2\right) \left(\sqrt{k^2+2 k l \cos \theta +l^2}+k\right)^3 \left(\sqrt{k^2+2 k l \cos \theta+l^2}+k+l\right)^4}\\
  &- \frac{2\pi l^4 \left(3 \sqrt{k^2+2 k l \cos \theta+l^2}+2 l\right)}{4 \left(k^2+2 k l \cos \theta +l^2\right) \left(\sqrt{k^2+2 k l \cos \theta +l^2}+k\right)^3 \left(\sqrt{k^2+2 k l \cos \theta+l^2}+k+l\right)^4}~.
  \end{split}
\end{align}
 The generic terms in the expansion \eqref{pertse} can be expressed using Apple Hypergeometric functions. However, to maintain brevity, we have omitted writing them explicitly.

The full amplitude at $ O(k_3^0)$ and $O(k_3^1)$ for $\phi^{4}$ theory in $D=d+1=4$, after loop integration, are,
\begin{eqn}
\left(\bm A^{(1)}_{3}\right)_0= 0.03125\frac{2\pi}{E_T}\frac{1}{k^2}\,,\quad \left(\bm A^{(1)}_{3}\right)_1 = -0.786073\frac{1}{E_T}  \frac{ k_3}{k^3}~. 
\end{eqn}
\vspace{0.1cm}

\section{CFT Correlators in Momentum Space}\label{app:CFTmom}
In this appendix, we review the derivation of the CFT correlation functions in momentum space. These are obtained via a Fourier transform of the correlation functions in position space \cite{Bzowski:2013sza}.
\subsection{2-point Function}
The 2-point function of scalar operators in the vacuum state in position space is given as 
\be
\braket{\mathcal{O}(\bm x)\mathcal{O}(0)}=\frac{C_{\mathcal O}}{x^{2\Delta}}~,
\ee
where $C_{\mathcal O}$ is a constant. Since the $d$-dimensional CFT is Euclidean we do not encounter any branch cuts while performing the Fourier transform. Thus, the integral that we have to evaluate is 
\be\label{cft2ptmom-1}
\braket{\braket{\mathcal O(\bm k) \mathcal O(-\bm k)}}=\int d^{d}x \frac{e^{-i \bm k\cdot \bm x}}{x^{2\Delta}}~,
\ee
where we use the notation $\braket{\braket{X}}$ to denote the correlation function without the momentum-conserving delta function. As the integral in equation \eqref{cft2ptmom-1} is axisymmetric, it can be evaluated in a similar method as the loop integrals in the main text,
\begin{eqn}
    \braket{\braket{\mathcal O(\bm k) \mathcal O(-\bm k)}}& =\int d^{d}x \frac{e^{-i \bm k\cdot\bm x}}{x^{2\Delta}} = S_{d-2} \int_{0}^{\infty} \frac{dr}{r^{2\Delta}} r^{d-1} \int_{0}^{\pi} d\theta \sin^{d-2}\theta e^{-ip r \cos\theta}~.
\end{eqn}
This integral can be expressed in terms of the Regularized Hypergeometric function ${}_0 \tilde F_{1}$\footnote{\bes
\int_0^\pi d\q \, \sin^b\q e^{- a \cos\q} = \sqrt{\pi} \Gamma\big[ \frac{1+b}{2} \big] \,{}_0 \tilde F_1 \Big[ 1 + \frac{ b}{2};  \frac{a^2}{4} \Big]
\ees}, 
\begin{eqn}
    \braket{\braket{\mathcal O(\bm k) \mathcal O(-\bm k)}}& = \sqrt{\pi} \Gamma\big[ \frac{d-1}{2} \big] S_{d-2}\int_{0}^{\infty} dr\, r^{d-2\Delta-1}  \,{}_0 \tilde F_1 \Big[ \frac{d}{2}; - \frac{k^2 r^2}{4} \Big] \,.
\end{eqn}
Evaluating this integral and plugging in the value of $S_{d-2}$ gives us 
\begin{eqn}
      \braket{\mathcal O(\bm k)  \mathcal O(-\bm k)}& = \pi^{d/2} 2^{d-2\Delta} k^{2\Delta-d}\,  \frac{\Gamma(\frac{d}{2}-\Delta)}{\Gamma(\Delta) }\, ,
\end{eqn}
which agrees with the conventional result for the momentum space two-point function in CFT$_d$ derived using Mellins-Barnes representation \cite{Bzowski:2013sza}. For $2\Delta - d = -1 \implies \Delta = 1$ in $d =3$ or $\Delta = 2$ in $d = 5$, which agrees with the results derived using the Witten diagrams.

\subsection{3-point Function}\label{app:3pt}
We do not state the derivation of the CFT 3-point function but just quote the result from \cite{Bzowski:2018fql}, which is expressed in terms of integrals over 3 Bessel-K functions
\begin{eqn}
\braket{\braket{\mathcal O_{\Delta_1}(\bm p_1) \mathcal O_{\Delta_2}(\bm p_2)\mathcal O_{\Delta_3}(\bm p_3)}} &= p_1^{\beta_1} p_2^{\beta_2} p_3^{\beta_3} \intsinf dx\, x^{\frac{d}{2}-1} K_{\beta_1}(p_1 x) K_{\beta_2}(p_2 x) K_{\beta_3}(p_3 x)
\end{eqn}
with $\beta_j = \Delta_j - \frac{d}{2}$. For the classes of Witten diagrams considered in this paper, we have $\Delta_j \equiv \Delta = \frac{d-1}{2}$, $d = 5$ and hence $\beta = - \frac{1}{2}$. This implies that 
\begin{eqn}\label{CFT3pt}
\braket{\braket{\mathcal O_{\Delta}(\bm p_1) \mathcal O_{\Delta}(\bm p_2)\mathcal O_{\Delta}(\bm p_3)}} &= \frac{1}{\sqrt{p_1 p_2 p_3}} \intsinf dx\ x^{\frac{d}{2} - 1} K_{-1/2}(p_1 x) K_{-1/2}(p_2 x) K_{-1/2}(p_3 x) \\
&= \frac{\pi^{3/2}}{2 \sqrt{2} (p_1 p_2 p_3) (p_1 + p_2 + p_3)}~.
\end{eqn}
This result agrees with the tree-level result of the three-point function computed using Witten diagrams. However, to make contact with the loop level diagram, i.e., the triangle diagram given in section \ref{sec:triangle}, we need to understand the behavior of this correlator for $\Delta \to 2$. This dependence is difficult to extract in general but it is possible to do it in the linearized order by using the integral representation for the Bessel functions. The functions that appear after performing the integrals contain $\log\frac{p_1 + p_2 + p_3}{p_3}$ and their permutations. Since the renormalized three-point diagram should match with equation \eqref{CFT3pt}, the analytic structure of the triangle diagram can be completely understood for the case of $\Delta = 2$ in $d = 5$ by studying the deviations away from $\Delta = 2$ and using the result of the diagram in the squeezed limit (given in \eqref{triangle-squeezed}).

\subsection{Relation with Cosmological Correlators in dS and Anomalous Dimensions}\label{sec:anomalousdimensions}
In this appendix, we demonstrate how one can use the wave functionals derived in this paper to compute cosmological correlation functions\cite{baumann-joyce}. For $\phi^4$ theory in four dimensions, the wave functions derived in this paper have the following expansion,
\begin{equation}\label{WF1}
\Psi[\phi] = e^{- \int d^3 k A(\bm k) \phi(\bm k) \phi(-\bm k) - \int d^3 k_1 d^3 k_2 d^3 k_3  B(\bm k_1, \bm k_2, \bm k_3) \phi(\bm k_1)\phi(\bm k_2) \phi(\bm k_3)\phi(-\bm k_1 - \bm k_2 - \bm k_3)}
\end{equation}
where the contributions from the loops to $A(\bm k)$ arise from equations \eqref{tadpole1}, \eqref{tadpole21}, \eqref{tadpole22}, \eqref{sunset} and similarly for $B(\bm k_1, \bm k_2, \bm k_3)$ in equation \eqref{4ptoneloop}\footnote{The tree level contributions are not given in our paper but can be found in \cite{Arkani-Hamed:2017fdk}. A similar analysis can be repeated for $\phi^3$ theory for which we have computed the 2 and 3-point functions.}. In order to avoid a clutter of notations, we schematically denote equation \eqref{WF1} as
\begin{equation}\label{WF2}
\Psi[\phi] = e^{- A \Phi^2 - B \Phi^4}~.
\end{equation}
The equal (and late time) correlation function $\braket{\phi(\bm k_1) \phi(\bm k_2) \cdots \phi(\bm k_n)}$ can be expressed in terms of the wave function as 
\begin{equation}\label{corr}
\braket{\phi(\bm k_1) \phi(\bm k_2) \cdots \phi(\bm k_n)} = \int D\phi \  |\Psi[\phi]|^2  \phi(\bm k_1) \phi(\bm k_2) \cdots \phi(\bm k_n)
\end{equation}
where we have suppressed the dependence on the time coordinate. As a concrete example, we shall demonstrate the computation for $n =4$. The functions $A(\bm k)$ and $B(\bm k_1, \bm k_2, \bm k_3)$ have the following perturbative expansions in the coupling constant, 
\begin{equation}
A = A_0 + \lambda A_1 + \lambda^2 A_2+ \cdots, \qquad 
B = \lambda B_1 + \lambda^2 B_2 +  \cdots~,
\end{equation}
where the important distinction is that $B$ always starts at $O(\lambda)$ whereas $A$ starts at $O(\lambda^0)$~. Therefore we can perform the path integral in \eqref{corr} in the same way we do in quantum mechanics. For this we can write $|\Psi[\phi]|^2$ in a more convenient form by expanding the exponential $e^{- B \Phi^4}$ perturbatively,
\begin{equation}
|\Psi[\phi]|^2 = e^{-2 \mbox{Re}A \Phi^2} \Big( 1 - \lambda \mbox{Re} B_1 \Phi^4 + \cdots\Big)
\end{equation}
This can be interpreted as the probability distribution for the path integral in equation \eqref{corr}. The 4-pt correlator is now given as,
\begin{equation}
\braket{\phi(\bm k_1)\cdots \phi(\bm k_4)}  = \int D \phi e^{-2 \mbox{Re}A \Phi^2} \Big( 1 - \lambda \mbox{Re} B_1 \Phi^4 + \cdots\Big)  \phi(\bm k_1)  \cdots \phi(\bm k_4)
\end{equation}
For demonstration, we focus on the two terms above. The first term is fairly simple and is given by Wick theorem,
\begin{equation}
 \int D \phi e^{-2 \mbox{Re}A \Phi^2}   \phi(\bm k_1) \cdots \phi(\bm k_4)\\
 = \braket{\phi(\bm k_1) \phi(\bm k_2)}  \braket{\phi(\bm k_3) \phi(\bm k_4)}  + \mbox{permutations}
\end{equation}
Where the functions $\braket{\phi(\bm q_1) \phi(\bm q_2)}$ are given as 
\begin{equation}
\braket{\phi(\bm q_1) \phi(\bm q_2)} = \frac{\delta^3(\bm q_1 + \bm q_2)}{2 \mbox{Re}A(\bm q_1)}
\end{equation}

The second term when written more explicitly is given as,
\begin{align}
&\lambda \int D \phi e^{-2 \mbox{Re}A \Phi^2}  \mbox{Re} B_1 \Phi^4 \phi(\bm k_1) \cdots \phi(\bm k_4) \nonumber \\
&= \lambda \int d^3 q_1 d^3 q_2 d^3 q_3 \mbox{Re} B_1(\bm q_1, \bm q_2, \bm q_3) \int D\phi \phi(\bm k_1) \cdots \phi(\bm k_4) \phi(\bm q_1) \phi(\bm q_2) \phi(\bm q_3) \phi(-\bm q_1 )  \\
&= \lambda \int d^3 q_1 d^3 q_2 d^3 q_3 \mbox{Re} B_1(\bm q_1, \bm q_2, \bm q_3) \braket{\phi(\bm k_1) \phi(\bm q_1)} \braket{\phi(\bm k_2) \phi(\bm q_2)} \braket{\phi(\bm k_3) \phi(\bm q_3)} \braket{\phi(\bm k_4) \phi(\bm q_4)} \nno \\
&\qquad + \mbox{permutations} \nno
\end{align}
where we have not neglected contributions from the disconnected pieces. This contributes to the four-point cosmological correlation function in terms of the wave function coefficients computed in the text.

 It is interesting to note that the 2-pt correlation function takes exactly the same form as that of a CFT in one lower spacetime dimension (discussed in the previous section) but depending on the renormalization scheme, it picks up an anomalous dimension. This is easily seen by expanding the 2-point function in the CFT about $\Delta = \frac{D-2}{2}$. For comparing with the results in our paper, we specifically consider the case when $D = 4$ and $D= 6$\footnote{In order to fix the normalization between the boundary operator $O(\bm k)$ and the bulk operator $\phi(\eta, \bm k)$ one can consider the bulk-bulk two-point function and extract the points onto the boundary. This is effectively the bulk-bulk propagator with both endpoints going to the boundary, i.e., $\lim_{\eta \to 0} G(\eta, \eta, \bm k)$. },
\begin{eqn}\label{anomalous}
\braket{\mathcal O(\bm k)  \mathcal O(-\bm k )}_{D = 4, \Delta = 1+\e} &= \frac{2 \pi ^2}{k}+ \frac{2 \pi ^2 \epsilon}{k}  \Big[2 \log \frac{k}{2}+\gamma_E -\psi ^{(0)}\left(\frac{1}{2}\right)\Big], \\
\braket{\mathcal O(\bm k)  \mathcal O(-\bm k)}_{D = 6, \Delta = 2+\e} &=\frac{2 \pi ^3}{k}+ \frac{2 \pi ^3 \epsilon}{k} \Big[2 \log\frac{k}{2} +\gamma_E -1-\psi ^{(0)}\left(\frac{1}{2}\right)\Big]  ~,
\end{eqn}
where $\gamma_E$ is the Euler Gamma constant and $\psi^{(0)}(x)$ is the Poly Gamma function. As seen from the equation above, the terms proportional to $\log k$ (which are also present in the loop diagrams) lead to anomalous dimensions.

\end{appendix}
\bibliographystyle{JHEP}
\bibliography{references}

\end{document}